\documentclass[twocolumn]{aastex63}

\usepackage{graphicx,textcomp,fancyhdr,hyperref,xcolor,fontawesome}
\definecolor{linkcolor}{rgb}{0.1216,0.4667,0.7059}
\usepackage[caption=false]{subfig}
\usepackage{float}

\shorttitle{Asteroseismology and planet detection for $\iota$~Draconis}
\shortauthors{Michelle L. Hill}

\begin{document}

\title{Asteroseismology of iota Draconis and Discovery of an Additional Long-Period Companion}

\author[0000-0002-0139-4756]{Michelle L. Hill}
\affiliation{Department of Earth and Planetary Sciences, University of California, Riverside, CA 92521, USA}
\email{mhill012@ucr.edu}
  
\author[0000-0002-7084-0529]{Stephen R. Kane}
\affiliation{Department of Earth and Planetary Sciences, University of California, Riverside, CA 92521, USA}

\author[0000-0002-4588-5389]{Tiago L. Campante}
\affiliation{Instituto de Astrof\'{\i}sica e Ci\^{e}ncias do Espa\c{c}o, Universidade do Porto, Rua das Estrelas, 4150-762 Porto, Portugal}
\affiliation{Departamento de F\'{\i}sica e Astronomia, Faculdade de Ci\^{e}ncias da Universidade do Porto, Rua do Campo Alegre, s/n, 4169-007 Porto, Portugal}

\author[0000-0002-4860-7667]{Zhexing Li}
\affiliation{Department of Earth and Planetary Sciences, University of
  California, Riverside, CA 92521, USA}

\author[0000-0002-4297-5506]{Paul A. Dalba}
\altaffiliation{NSF Astronomy and Astrophysics Postdoctoral Fellow}
\affiliation{Department of Earth and Planetary Sciences, University of California, Riverside, CA 92521, USA}

\author[0000-0003-2630-8073]{Timothy D.~Brandt}
\affiliation{Department of Physics, University of California, Santa Barbara, Santa Barbara, CA 93106, USA}

\author[0000-0002-6980-3392]{Timothy R. White}
\affiliation{Sydney Institute for Astronomy (SIfA), School of Physics, University of Sydney, NSW 2006, Australia}

\author[0000-0003-2595-9114]{Benjamin J.S. Pope}
\affiliation{School of Mathematics and Physics, The University of Queensland, St Lucia, QLD 4072, Australia}
\affiliation{Centre for Astrophysics, University of Southern Queensland, West Street, Toowoomba, QLD 4350, Australia}

\author[0000-0002-3481-9052]{Keivan G. Stassun}
\affiliation{Vanderbilt University, Department of Physics \& Astronomy, 6301 Stevenson Center Lane, Nashville, TN 37235, USA}

\author[0000-0003-3504-5316]{Benjamin J. Fulton}
\affiliation{NASA Exoplanet Science Institute/Caltech-IPAC, MC 314-6, 1200 E California Blvd, Pasadena, CA 91125, USA}

\author[0000-0001-8835-2075]{Enrico Corsaro}
\affiliation{INAF --- Osservatorio Astrofisico di Catania, via S.~Sofia 78, 95123 Catania, Italy}

\author[0000-0001-6396-2563]{Tanda Li}
\affiliation{School of Physics and Astronomy, University of Birmingham, Edgbaston, Birmingham B15 2TT, UK}
\affiliation{Stellar Astrophysics Centre (SAC), Department of Physics and Astronomy, Aarhus University, Ny Munkegade 120, 8000 Aarhus C, Denmark}

\author[0000-0001-7664-648X]{J. M. Joel Ong}
\affiliation{Department of Astronomy, Yale University, 52 Hillhouse Ave., New Haven, CT 06511, USA}

\author[0000-0001-5222-4661]{Timothy R. Bedding}
\affiliation{Sydney Institute for Astronomy (SIfA), School of Physics, University of Sydney, NSW 2006, Australia}
\affiliation{Stellar Astrophysics Centre (SAC), Department of Physics and Astronomy, Aarhus University, Ny Munkegade 120, 8000 Aarhus C, Denmark}

\author[0000-0002-9480-8400]{Diego Bossini}
\affiliation{Instituto de Astrof\'{\i}sica e Ci\^{e}ncias do Espa\c{c}o, Universidade do Porto, Rua das Estrelas, 4150-762 Porto, Portugal}

\author[0000-0002-1988-143X]{Derek L. Buzasi}
\affiliation{Department of Chemistry and Physics, Florida Gulf Coast University, 10501 FGCU Blvd. S., Fort Myers, FL 33965, USA}

\author[0000-0002-5714-8618]{William J. Chaplin}
\affiliation{School of Physics and Astronomy, University of Birmingham, Edgbaston, Birmingham B15 2TT, UK}
\affiliation{Stellar Astrophysics Centre (SAC), Department of Physics and Astronomy, Aarhus University, Ny Munkegade 120, 8000 Aarhus C, Denmark}

\author[0000-0001-8237-7343]{Margarida S. Cunha}
\affiliation{Instituto de Astrof\'{\i}sica e Ci\^{e}ncias do Espa\c{c}o, Universidade do Porto,  Rua das Estrelas, 4150-762 Porto, Portugal}

\author[0000-0002-8854-3776]{Rafael A. Garc\'\i a}
\affiliation{AIM, CEA, CNRS, Universit\'e Paris-Saclay, Universit\'e Paris Diderot, Sorbonne Paris Cit\'e, F-91191 Gif-sur-Yvette, France}

\author[0000-0003-0377-0740]{Sylvain N. Breton}
\affiliation{AIM, CEA, CNRS, Universit\'e Paris-Saclay, Universit\'e Paris Diderot, Sorbonne Paris Cit\'e, F-91191 Gif-sur-Yvette, France}

\author[0000-0003-2400-6960]{Marc Hon}
\affiliation{Institute for Astronomy, University of Hawai`i, 2680 Woodlawn Drive, Honolulu, HI 96822, USA}
\affiliation{School of Physics, The University of New South Wales, Sydney NSW 2052, Australia}

\author[0000-0001-8832-4488]{Daniel Huber}
\affiliation{Institute for Astronomy, University of Hawai`i, 2680 Woodlawn Drive, Honolulu, HI 96822, USA}

\author[0000-0002-7614-1665]{Chen Jiang}
\affiliation{Max-Planck-Institut f\"ur Sonnensystemforschung, Justus-von-Liebig-Weg 3, 37077 G\"ottingen, Germany}
\affiliation{School of Physics and Astronomy, Sun Yat-Sen University, No. 135, Xingang Xi Road, Guangzhou 510275, P. R. China}

\author[0000-0001-9198-2289]{Cenk Kayhan}
\affiliation{Department of Astronomy and Space Sciences, Erciyes University, 38030, Kayseri, Turkey}

\author[0000-0002-3322-5279]{James S. Kuszlewicz}
\affiliation{Max-Planck-Institut f\"ur Sonnensystemforschung, Justus-von-Liebig-Weg 3, 37077 G\"ottingen, Germany}
\affiliation{Stellar Astrophysics Centre (SAC), Department of Physics and Astronomy, Aarhus University, Ny Munkegade 120, 8000 Aarhus C, Denmark}

\author[0000-0002-0129-0316]{Savita Mathur}
\affiliation{Instituto de Astrof\'isica de Canarias (IAC), E-38205 La Laguna, Tenerife, Spain}
\affiliation{Universidad de La Laguna (ULL), Departamento de Astrof\'isica, E-38206 La Laguna, Tenerife, Spain}

\author[0000-0001-6359-2769]{Aldo Serenelli}
\affiliation{Institute of Space Sciences (ICE, CSIC) Campus UAB, Carrer de Can Magrans, s/n, E-08193, Bellaterra, Spain}
\affiliation{Institut d'Estudis Espacials de Catalunya (IEEC), C/Gran Capit\`a, 2-4, E-08034, Barcelona, Spain}

\author[0000-0002-4879-3519]{Dennis Stello}
\affiliation{School of Physics, The University of New South Wales, Sydney NSW 2052, Australia}
\affiliation{Stellar Astrophysics Centre (SAC), Department of Physics and Astronomy, Aarhus University, Ny Munkegade 120, 8000 Aarhus C, Denmark}


\begin{abstract}

Giant stars as known exoplanet hosts are relatively rare due to the potential challenges in acquiring precision radial velocities and the small predicted transit depths. However, these giant host stars are also some of the brightest in the sky and so enable high signal-to-noise follow-up measurements. Here we report on new observations of the bright ($V \sim 3.3$) giant star $\iota$~Draconis ($\iota$~Dra), known to host a planet in a highly eccentric $\sim$511 day period orbit. {\it TESS} observations of the star over 137 days reveal asteroseismic signatures, allowing us to constrain the stellar radius, mass, and age to $\sim$2\,\%, $\sim$6\,\%, and $\sim$28\,\%, respectively. We present the results of continued radial velocity monitoring of the star using the Automated Planet Finder over several orbits of the planet. We provide more precise planet parameters of the known planet and, through the combination of our radial velocity measurements with Hipparcos and Gaia astrometry, we discover an additional long-period companion with an orbital period of $\sim68^{+60}_{-36}$~years. Mass predictions from our analysis place this sub-stellar companion on the border of the planet and brown dwarf regimes. The bright nature of the star combined with the revised orbital architecture of the system provides an opportunity to study planetary orbital dynamics that evolve as the star moves into the giant phase of its evolution.

\end{abstract}



\section{Introduction}
\label{intro}

Exoplanets have been discovered around a diversity of stellar types and with a broad range of orbital architectures \citep{ford2014,winn2015}. Despite challenges with regards to stellar pulsations \citep{hatzes2018}, radial velocity (RV) surveys for planets orbiting giant stars are underway \citep{hekker2007,reffert2015}. One of the brightest and nearest giant stars known to host a planet is $\iota$~Draconis (hereafter $\iota$~Dra); a $V = 3.29$ K2 giant star located at a distance of $\sim$31~pc. At the time of its detection by \citet{frink2002}, $\iota$~Dra~b, or HD~137759~b, was the first planet to be found to orbit a giant star. The initial detection, based on RV observations of a full planetary orbit, revealed an orbital period of $\sim536$ days, an eccentricity of 0.70, a minimum planet mass of 8.9~$M_J$, and a semi-major axis of 1.3~AU. These properties were refined by \citet{zechmeister2008} with the help of an increased observational RV baseline and resulted in the detection of an additional linear trend in the residuals to the single planet solution. The planet orbital properties were further refined by \citet{kane2010a} who confirmed the existence of the linear trend detection. To rule out the possibility of a stellar companion as the cause of the linear trend, \citet{kane2014c} investigated $\iota$~Dra with the Differential Speckle Survey Instrument (DSSI) on the Gemini North telescope \citep{horch2009}. These observations were able to exclude bound low-mass M-dwarfs at wide separations. As the long term linear trend continued, it was becoming increasingly likely that it was caused by a substellar companion.

The potential of a transit event of the $\iota$~Dra planet was evaluated in \citet{kane2010a}. As planets orbiting giant stars tend to have
large transit probabilities due to the size of the host stars \citep{assef2009} and planets with higher eccentricity also have an increased probability of transiting \citep{barnes2007d,kane2008b}, $\iota$~Dra~b is expected to have a relatively high transit probability of $\sim$16.5\% \citep{kane2010a}. In a recent study, \citet{dalba2019c} found $\iota$~Dra had a 11$\pm$3\% probability of having a transiting geometry, and subsequently a 1.8$\pm^{0.6}_{0.5}$\% probability of the Transiting Exoplanet Survey Satellite (TESS) seeing the transit in the primary mission.

Given the high dependence of transit probabilities and other planetary properties on the host star parameters, extracting reliable stellar properties is crucially important for continued studies of the system. Interferometric observations of $\iota$~Dra by \citet{baines2011} measured a stellar radius of $\sim$12~$R_\odot$ and an effective temperature of $T_\mathrm{eff} = 4545$~K. However, long-term and continuous precision photometry of the star is challenging due to its brightness and northern celestial location, restricting access from many facilities. Such a photometric dataset would not only allow the possibility of transit detection, but enable a concise evaluation of the stellar properties via asteroseismology, which is particularly well suited to giant stars due to the large amplitudes and accessible frequencies available \citep{campante2016b}.

Here we present a new analysis of the $\iota$~Dra system that includes a substantially updated RV dataset, Hipparcos and Gaia astrometry and precision photometry from TESS. In particular, the combination of astrometry with the new RV data demonstrate that the linear trend has finally revealed a curvature, allowing an orbital period for the second companion to be estimated. In Section~\ref{obs} we describe the new RV data, and our analysis of the photometry from TESS. We provide refined stellar properties in Section \ref{stellar}, including an SED analysis and an asteroseismic study enabled by the precision data from TESS. We present our revised orbital properties of $\iota$~Dra~b and of the long-term RV trend in Section~\ref{planet}, including a dynamical analysis with the \texttt{MEGNO} chaos indicator, orbital constraints on the outer companion from \texttt{The Joker}, and our best fit for the additional companion through the combination of RV and astrometry using \texttt{htof} and \texttt{orvara}. In Section~\ref{discussion} the orbital dynamics of planets in evolved systems are discussed and conclusions and future directions are provided in Section~\ref{conclusions}.


\section{Observations}
\label{obs}


\subsection{TESS Photometry}
\label{tess}

The TESS mission is designed to survey nearby F, G, K, and M type stars for signatures of transiting exoplanets \citep{ricker2015}. TESS observations of $\iota$~Dra occurred during Sectors 15, 16, 22, 23, 24 at 2-minute cadence. 
At a magnitude of $T_{\text{mag}} = 2.27$, $\iota$~Dra is significantly saturated as seen by the TESS detector, and accordingly required special processing to obtain a high-quality lightcurve. In particular, with the large postage stamp, spatially-varying background light is a major source of noise \citep{eisner19, dalba2020a}, and to ameliorate this we create a spatially-varying background model using a second-order polynomial fitted to pixels at the edge of the aperture. As well as this, the default aperture from the SPOC pipeline was too small, and we instead created our own: first we defined a threshold mask at 50\% maximum amplitude using \texttt{lightkurve} \citep{lightkurve}, and applied a binary dilation to expand this by one pixel in each direction. The Jupyter notebook used to generate this lightcurve is available on GitHub.  \href{https://github.com/hvidy/tessbkgd/blob/stable/notebooks/iot_Dra_tpf.ipynb}{\color{linkcolor}\faGithub}

Upon inspection of the TESS photometry we found no indications of any transiting planets. However, none of the TESS observations of $\iota$~Dra coincide with the expected time of conjunction for $\iota$~Dra~b, so a transit of this planet cannot be ruled out. Future observations of $\iota$~Dra by TESS may coincide with the time of inferior conjunction of $\iota$~Dra~b. This is discussed further in Section \ref{discussion}.


\subsection{Radial Velocities}
\label{rvs}

A total of 165 RV observations obtained by the 0.6~m Coud\'{e} Auxiliary Telescope (CAT) and the Hamilton \'{E}chelle Spectrograph (HES) at the Lick Observatory in California were extracted from previous published works by  \citet{frink2002,butler2006,zechmeister2008,kane2010a}. 

An additional 456 RV observations were obtained by the Levy
spectrometer on the Automated Planet Finder (APF) \citep{radovan2014,vogt2014a} at Lick Observatory between February 2018 to February 2021.  
The spectra were reduced using the standard procedures of the
California Planet Search \citep{Howard_2010}. A subset of the APF RV dataset is found in Table \ref{APF}. The full dataset will be made available in machine-readable form.

\floattable
\startlongtable
\begin{deluxetable}{llll}
  \tablewidth{0pc}
 \tabletypesize{\scriptsize}
  \centering
  \tablecaption{Radial Velocity Observations of $\iota$~Dra.\tablenotemark{a}
  \label{APF}}
\tablehead{
\colhead{Time}  & \colhead{RV} & \colhead{Uncertainty} &  \colhead{Telescope}  \\
\colhead{BJD}  & \colhead{m/s} & \colhead{m/s} &  \colhead{}}
\startdata
2458156.925 & -194.377146 & 2.292272 & apf \\
2458156.925 & -202.796987 & 2.548035 & apf \\
2458156.926 & -196.984888 & 2.431443 & apf \\
2458160.96 & -229.261174 & 2.014732 & apf \\
2458160.96 & -226.650269 & 2.191855 & apf \\
2458160.961 & -229.033612 & 2.148611 & apf \\
2458161.065 & -217.703627 & 2.105308 & apf \\
2458161.065 & -219.498013 & 3.215146 & apf \\
2458161.066 & -218.682983 & 2.15082 & apf \\
2458162.943 & -231.747948 & 2.204404 & apf \\
\enddata
\tablenotetext{a}{This Table is a subset of the full dataset which will be made available in machine-readable form.}
\end{deluxetable}


\section{Stellar Characterization}
\label{stellar}


\subsection{Spectral Energy Distribution}\label{sec:sed}

\begin{figure}[!ht]
\centering
\includegraphics[width=2.5in,angle=90,trim=75 80 85 110,clip]{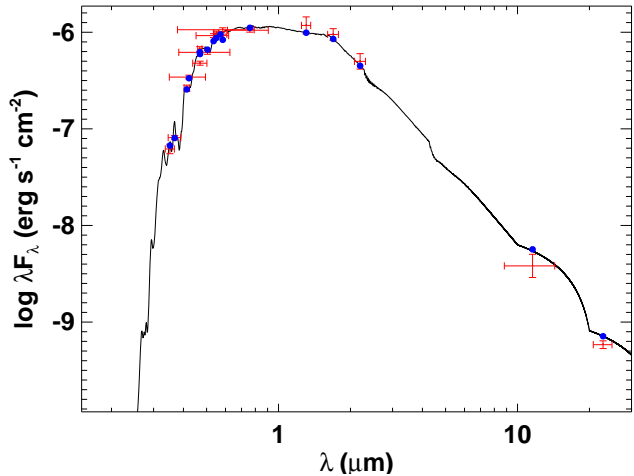}
\caption{Spectral energy distribution of $\iota$~Dra. Red symbols represent the observed photometric measurements, where the horizontal bars represent the effective width of the passband. Blue symbols are the model fluxes from the best-fit Kurucz atmosphere model (black).  \label{fig:sed}}
\end{figure}

As an independent determination of the basic stellar parameters, we performed an analysis of the broadband spectral energy distribution (SED) of the star together with the {\it Gaia\/} DR2 parallaxes \citep[adjusted by $+0.08$~mas to account for the systematic offset reported by][]{StassunTorres:2018}, in order to determine an empirical measurement of the stellar radius, following the procedures described in \citet{Stassun:2016,Stassun:2017,Stassun:2018}. We pulled the $UBV$ magnitudes from \citet{Mermilliod:1991}, the $uvby$ Str\"omgren magnitudes from \citet{Paunzen:2015}, the $B_T V_T$ magnitudes from {\it Tycho-2}, the $JHK_S$ magnitudes from {\it 2MASS}, the W3--W4 magnitudes from {\it WISE}, and the $G G_{\rm BP} G_{\rm RP}$ magnitudes from {\it Gaia}. Together, the available photometry spans the full stellar SED over the wavelength range 0.3--22~$\mu$m (see Figure~\ref{fig:sed}).

We performed a fit using Kurucz stellar atmosphere models, with the effective temperature ($T_{\rm eff}$), metallicity ([Fe/H]), and surface gravity ($\log g$) adopted from the spectroscopic analysis of \citet{Jofre:2015}. The only additional free parameter is the extinction ($A_V$), which we set to zero given the star's proximity. The resulting fit is very good (Figure~\ref{fig:sed}) with a reduced $\chi^2$ of 2.8. Integrating the (unreddened) model SED gives the bolometric flux at Earth, $F_{\rm bol}~=~1.692~\pm~0.059~\times~10^{-6}$~erg~s$^{-1}$~cm$^{-2}$. Taking the $F_{\rm bol}$ and $T_{\rm eff}$ together with the {\it Gaia\/} DR2 parallax, gives the stellar radius, $R_\star~=~11.94~\pm~0.32~R_\odot$. In addition, we can use the $R_\star$ together with the spectroscopic $\log~g$ to obtain an empirical mass estimate of $M_\star~=~1.72~\pm~0.29~M_\odot$, which is roughly consistent with that estimated via the eclipsing-binary based empirical relations of \citet{torres2010}, $M_\star~=~2.23~\pm~0.13$~M$_\odot$. Finally, from the spectroscopic $v\sin i$ together with $R_\star$ we obtain an estimate of the stellar rotation period lower limit, $P_{\rm rot}/\sin~i~=~325~\pm~78$~d. 


\subsection{Asteroseismology}


\subsubsection{Global Oscillation Parameters}
\begin{figure*}[!t]
\centering
\includegraphics[width=\linewidth]{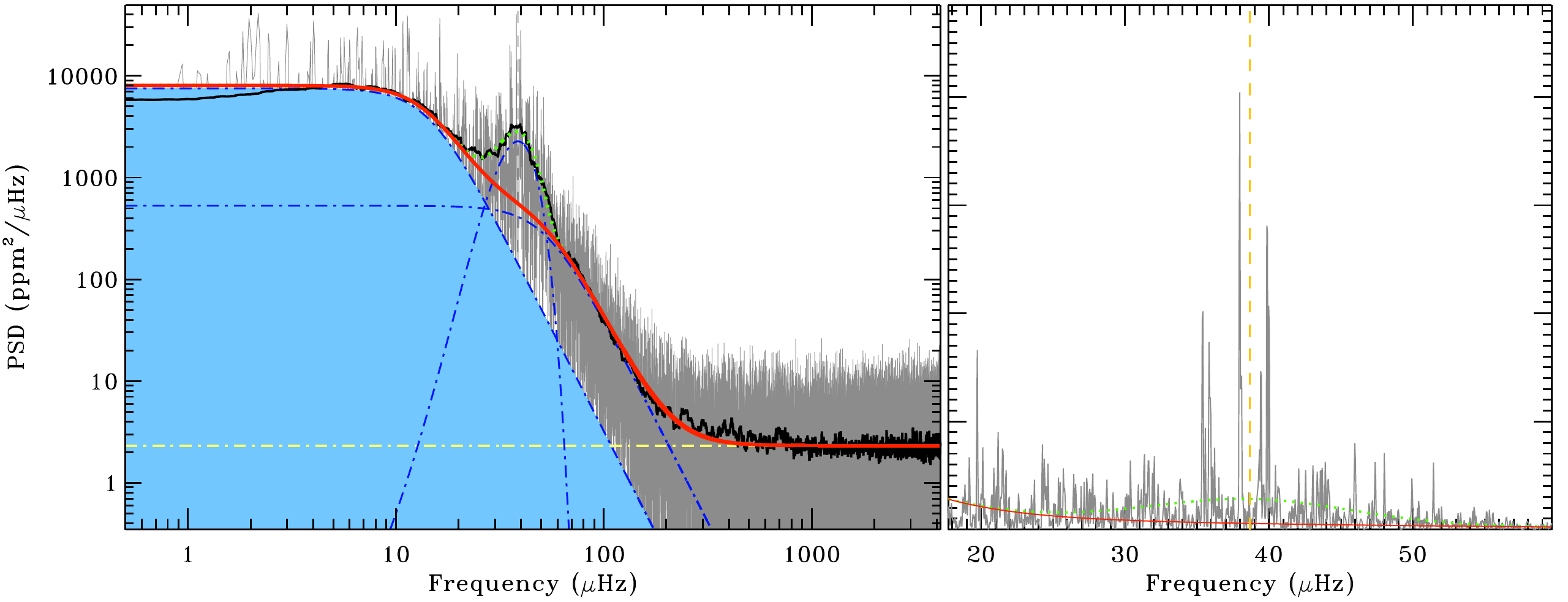}
\caption{Power spectrum of $\iota$~Dra based on the full {\it TESS} light curve. Left Panel: The power spectrum is shown in gray on a logarithmic scale (with a heavily smoothed version in black). The solid red curve is a fit to the background, consisting of two Harvey-like profiles (blue dot-dashed curves) plus white noise (yellow, horizontal dot-dashed line). A joint fit to the oscillation power excess (blue dot-dashed Gaussian curve) and background is visible at $\sim40\:{\rm \mu Hz}$ as a dotted green curve. Right Panel: The power spectrum is shown on a linear scale and centered on the oscillations. The vertical yellow dashed line is a proxy for $\nu_{\rm max}$. The remaining curves have the same meaning as in the left panel.\label{fig:PSD}}
\end{figure*}

Figure \ref{fig:PSD} shows the power spectrum of $\iota$~Dra based on the full {\it TESS} light curve extracted in Sect.~\ref{tess}. It reveals a clear power excess due to solar-like oscillations at $\sim40\:\rm{\mu Hz}$. This is in agreement with \citet{zechmeister2008}, who measured solar-like oscillations with frequencies around 34.7-46.3 $\mu$Hz.

We started by measuring the large frequency separation, $\Delta\nu$, and the frequency of maximum oscillation amplitude, $\nu_{\rm max}$, using a range of well-tested automated analysis methods \citep{huber2009,mosser2009,mathur2010,DIAMONDS,campante2017,campante2019,montellano2018,lightkurve,viani2019,famed}, which have previously been extensively applied to {\it Kepler}/{\it K2} data. Returned values were subject to a preliminary step which involved the rejection of outliers following Peirce's criterion \citep{peirce1852,gould1855}. A final, consolidated pair of values, $\Delta\nu = 4.02 \pm 0.02 \: \rm{\mu Hz}$ and $\nu_{\rm max} = 38.4 \pm 0.5 \: \rm{\mu Hz}$, then stem from the source/method \citep{huber2009} which minimizes the normalized rms deviation about the median. Uncertainties are the corresponding formal uncertainties.

\subsubsection{Individual Mode Frequencies}
A total of $N=7$ fitters extracted individual mode frequencies from the power spectrum. Methods employed ranged from an iterative sine-wave fitting approach \citep[e.g.,][]{period04,bedding2010} to the fitting of Lorentzian and sinc$^2$ mode profiles \citep[e.g.,][]{HandCamp11,montellano2018,famed}. We then followed the procedure described in \citet{campante2011} to produce two frequency lists, namely, a {\it minimal frequency list} and a {\it maximal frequency list}. The former includes modes (after outlier rejection) detected by more than $\lfloor N/2\rfloor$ fitters. One may think of it as a conservative list (16 modes). The latter includes modes (after outlier rejection) detected by at least 2 fitters (31 modes). The minimal list is thus a subset of the maximal list. Only those modes belonging to the minimal list will be effectively modeled in Sect.~\ref{sec:detailmodel}. Hereafter, we adopt a set of observed mode frequencies (and corresponding uncertainties) tracing back to a single fitter/method, namely, \texttt{FAMED} \citep{famed}, so as to guarantee reproducibility. Table~\ref{tab:modes} lists all significant modes (i.e., $p_{\rm det} \ge 0.993$, see Section 5.3 of \citep{famed} for details) returned by \texttt{FAMED} (note that not all modes belonging exclusively to the maximal list were found to be significant by \texttt{FAMED} and have thus not been listed). The dominant mode found by \citet{zechmeister2008} coincides with the first mode listed in Table~\ref{tab:modes}.


\subsubsection{Evolutionary State}\label{sec:evol}
Measurement of the period spacing between mixed modes allows distinguishing between hydrogen-shell burning (or red-giant branch; RGB) and helium-core burning (HeB) red giants \citep{bedding2011,mosser2011}. Estimation of the period spacing, $\Delta\Pi_1$, is, however, not possible when considering modes in the minimal list, owing to the limited number of observed dipole mixed modes per radial order. Having further run the \citet{vrard2016} method and evaluated the asymptotic acoustic-mode offset \citep{kallinger2012,jcd2014} also proved inconclusive.

We next resorted to machine learning classification methods. We employed the deep learning method of \citet{hon2017,hon2018}, which efficiently classifies the evolutionary state of oscillating red giants by recognizing visual features in their power spectra. Application of this method points towards an RGB classification with high confidence ($p > 0.9$). Alternatively, we made use of the \texttt{Clumpiness} evolutionary state classifier \citep{clumpiness}, which returned a probability of $\sim 0.8$ (when applied to the full {\it TESS} light curve) of the star being on the RGB.

\subsubsection{Detailed Stellar Modeling}\label{sec:detailmodel}
We modeled the modes in the minimal list, together with a set of classical constraints (namely, $T_{\rm eff}$, $[{\rm Fe}/{\rm H}]$, and $L_\ast$; see Table~\ref{tab:starprop}), following the methodology of \citet{li2020}, without considering interpolation and setting the model systematic uncertainty to zero. The underlying grid of stellar models is described in Appendix \ref{APPGrid}. Figure \ref{fig:echelle} is an \'echelle diagram showing the frequency match for a representative best-fitting model in the grid. We note that only RGB models were able to provide a sensible fit to the observed frequencies within the quoted $T_{\rm eff}$, $[{\rm Fe}/{\rm H}]$, and $L_\ast$ ranges (we used 5$\sigma$ ranges). This constitutes further evidence in support of the RGB classification.

We provide values from detailed modeling for the stellar mass ($M_\ast$), radius ($R_\ast$), surface gravity ($\log g$), and age ($t$) in Table \ref{tab:starprop}. Quoted uncertainties include both a statistical and a systematic contribution. The latter accounts for the impact of using different model grids --- covering a range of input physics --- and analysis methodologies on the final estimates, full details of which will be presented in a follow-up paper (T.~Campante et al.~2021, in preparation). We note the excellent agreement (within 1 $\sigma$) between the seismic and interferometric radii.

\begin{figure}[!t]
\centering
\includegraphics[width=\linewidth]{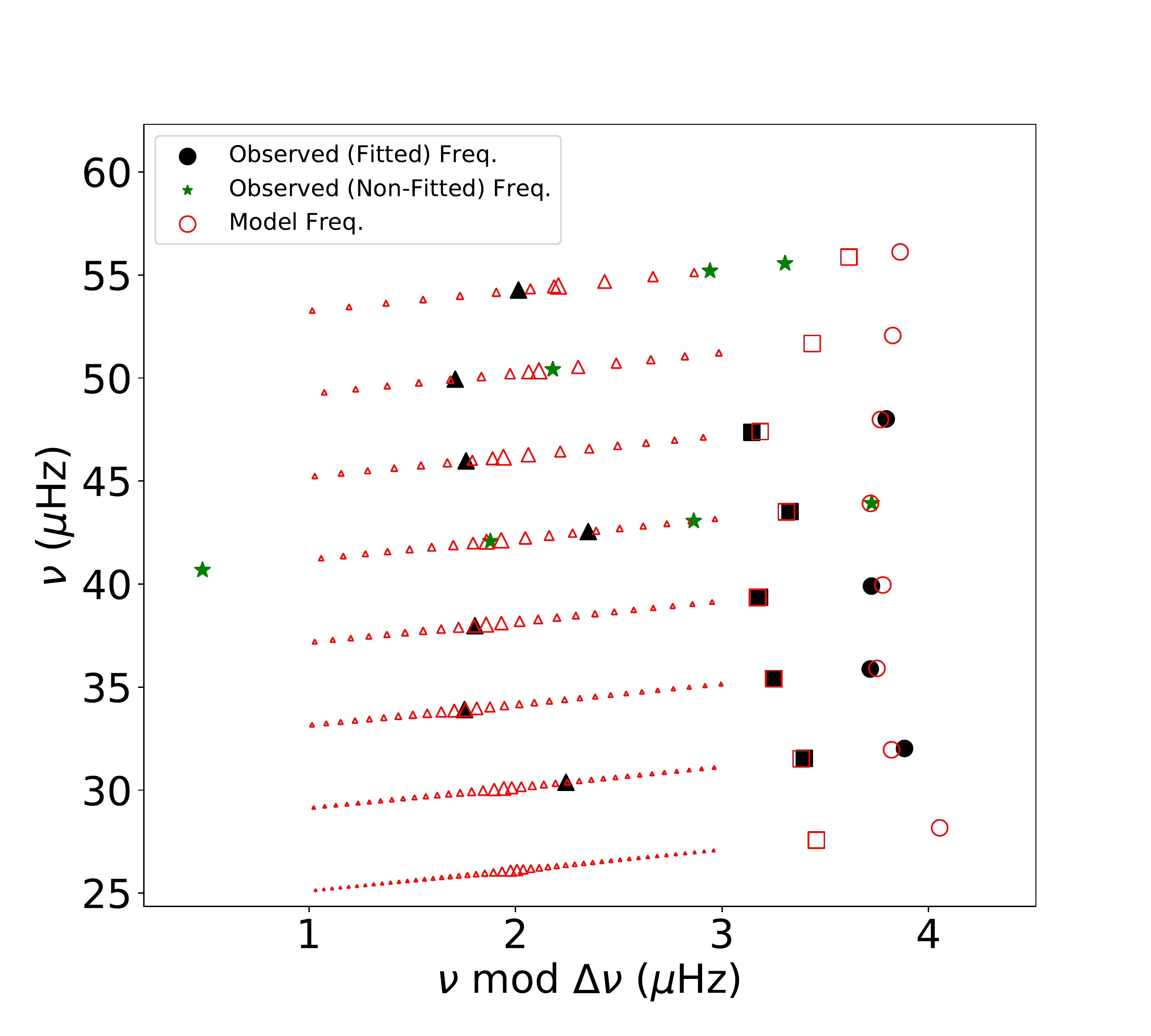}
\caption{\'Echelle diagram displaying the observed frequencies as well as the model frequencies corresponding to a representative best-fitting model (no interpolation was used and so this corresponds to a specific model in the grid). Filled black symbols represent observed (fitted) frequencies, i.e., belonging to the minimal list. Open red symbols are model frequencies. Circles, triangles, and squares indicate modes of angular degree $\ell=0$ (radial modes), $\ell=1$ (dipole modes), and $\ell=2$ (quadrupole modes), respectively. Green five-pointed stars correspond to observed frequencies not in the minimal list. The extra $\ell=2$ model frequencies correspond to the most p-like modes, whereas we have plotted a range of g-dominated $\ell=1$ model frequencies per order (with symbol size scaled as $I^{-0.5}$, $I$ being the mode inertia).\label{fig:echelle}}
\end{figure}

\begin{table}[!t]
\begin{center}
\caption{Stellar Parameters.\label{tab:starprop}}
\renewcommand{\tabcolsep}{0mm}
\begin{tabular}{l c c}
\noalign{\smallskip}
\tableline\tableline
\noalign{\smallskip}
\textbf{Parameter} & \textbf{Value} & \textbf{Source} \\
\noalign{\smallskip}
\tableline
\noalign{\smallskip}
\multicolumn{3}{c}{Basic Properties} \\
\noalign{\smallskip}
\hline
\noalign{\smallskip}
{\it Gaia} ID & DR2 1614731957530945280 & 1 \\
TIC & 165722603 & 2 \\
{\it TESS} Mag. & 2.27 & 2 \\
Sp.~Type & K2 III & 3 \\
\noalign{\smallskip}
\hline
\noalign{\smallskip}
\multicolumn{3}{c}{Spectroscopy} \\
\noalign{\smallskip}
\hline
\noalign{\smallskip}
$T_{\rm eff}$ (K) & $4504 \pm 62$\tablenotemark{a} & 4 \\
$[{\rm Fe}/{\rm H}]$ (dex) & $0.03 \pm 0.08$\tablenotemark{a} & 4 \\
$\log g$ (cgs) & $2.52 \pm 0.07$ & 4 \\
\noalign{\smallskip}
\hline
\noalign{\smallskip}
\multicolumn{3}{c}{SED \& \textit{Gaia} DR2 Parallax} \\
\noalign{\smallskip}
\hline
\noalign{\smallskip}
$F_{\rm bol}$ (${\rm erg\,s^{-1}\,cm^{-2}}$) & $(1.692 \pm 0.059) \times 10^{-6}$ & 5 \\
$R_\ast$ (${\rm R}_\odot$) & $11.94 \pm 0.32$ & 5 \\
$L_\ast$ (${\rm L}_\odot$) & $52.78 \pm 2.10$\tablenotemark{b} & 5 \\
$\pi$ (mas) & $31.65 \pm 0.30$\tablenotemark{c} & 1 \\
\noalign{\smallskip}
\hline
\noalign{\smallskip}
\multicolumn{3}{c}{Asteroseismology} \\
\noalign{\smallskip}
\hline
\noalign{\smallskip}
$\Delta\nu$ ($\mu$Hz) & $4.02 \pm 0.02$ & 5 \\
$\nu_{\rm max}$ ($\mu$Hz) & $38.4 \pm 0.5$ & 5 \\
$M_\ast$ (${\rm M}_\odot$) & $1.54\pm0.09$\tablenotemark{d} & 5 \\
$R_\ast$ (${\rm R}_\odot$) & $11.79\pm0.24$\tablenotemark{d} & 5 \\
$\log g$ (cgs) & $2.48\pm0.01$\tablenotemark{d} & 5 \\
$t$ (Gyr) & $2.65\pm0.74$\tablenotemark{d} & 5 \\
\noalign{\smallskip}
\hline
\noalign{\smallskip}
\end{tabular}
\end{center}
\tablenotetext{a}{\scriptsize Formal uncertainties have been inflated according to \citet{torres2012}.}
\tablenotetext{b}{\scriptsize Based on SED fit (Sect.~\ref{sec:sed}) and {\it Gaia} DR2 parallax.}
\tablenotetext{c}{\scriptsize Adjusted for the systematic offset of \citet{StassunTorres:2018}.}
\tablenotetext{d}{\scriptsize Uncertainties include both a statistical and a systematic contribution (added in quadrature).}
\tablerefs{\scriptsize (1) \citet{GaiaDR2}, (2) \citet{TIC}, (3) \citet{spectype}, (4) \citet{Jofre:2015}, (5) this work.}
\end{table}


\section{Detection of a Long-Period Companion}
\label{planet}

\subsection{Radial Velocity Analysis}
\label{RV_analysis}

\begin{figure*}
\centering
\includegraphics[width=0.7\textwidth]{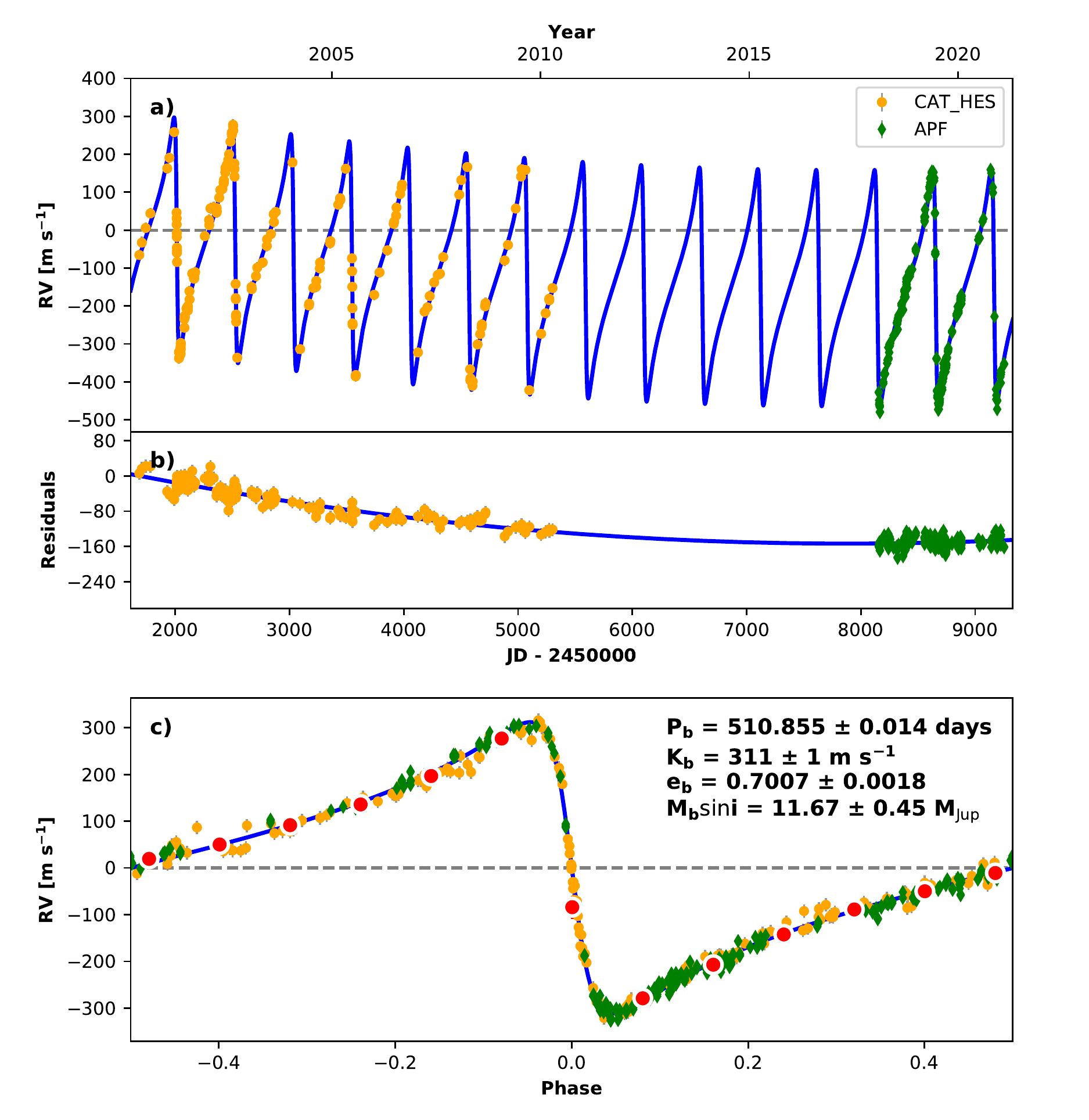}
\caption{Best-fit single-planet Keplerian orbital model for $\iota$~Dra. Yellow data points were taken using Coud\'{e} Auxiliary Telescope with the Hamilton \'{E}chelle Spectrograph (CAT\_HES) and green were taken using the Automated Planet Finder (APF) with the Levy spectrometer. The maximum likelihood model is plotted.  The thin blue line is the best-fit model.  {\bf b)} Residuals to the best-fit 1-planet model. The curvature of the residuals have flattened out, indicating that this system is turning around. {\bf c)} RVs phase-folded to the ephemeris of planet b. The small point colors and symbols are the same as in panel {\bf a}.  Red circles are the same velocities binned in 0.08 units of orbital phase.  The phase-folded model for planet b is shown as the blue line. See Table~\ref{tab:Radvel2planet} for the definition of the orbital parameters shown in panel {\bf c}. }
\label{fig:137759}
\end{figure*}

The RV data for $\iota$~Dra were fit using the RV modeling toolkit \texttt{RadVel} \citep{fulton2018a} in order to refine the orbital solution and look for curvature within the previously reported linear trend to determine if there were indications for additional planetary companions.
\texttt{RadVel} enables users to model Keplerian orbits in radial velocity time series. \texttt{RadVel} fits RVs using an iterative approach to solve the set of equations for
the Keplerian orbit to
determine the best fit for the observed RV curve. It then employs modern Markov chain Monte Carlo (MCMC) sampling techniques \citep{metropolis53,hastings70,Foreman2013} and robust convergence criteria to ensure accurately estimated orbital parameters and their associated uncertainties.
Once the MCMC chains are well mixed, \texttt{RadVel} then supplies an output of the final parameter values from the Maximum A Posteriori (MAP) fit.

We used the previously published orbital values from \citet{butler2006} as priors for $\iota$~Dra~b and allowed all orbital parameters, including the linear and curvature terms, to be free. The best-fit solution from \texttt{RadVel} gave $\iota$~Dra~b an orbital period of 510.855~$\pm$~0.014~days, a semi-amplitude of 311~$\pm$~1~$ms^{-1}$, eccentricity 0.7008~$\pm$~0.0018 and using our $M_*$ value from Table~\ref{tab:starprop}, a derived $M_p\sin i$~of~11.67~$\pm$~0.45~$M_{J}$. The central 68\% confidence intervals computed from the MCMC chains are presented in Table~\ref{tab:Radvel2planet}. 
The preferred model includes linear and curvature terms for $\iota$~Dra. This indicates an additional body in orbit around $\iota$~Dra. The residuals of the single planet model can be seen to flatten out, indicating the additional orbiting body has reached quadrature (Figure \ref{fig:137759}).

\begin{deluxetable*}{lrr}[ht]
\tablecaption{\label{tab:Radvel2planet} RadVel MCMC Posteriors for $\iota$~Dra~b}
\tablehead{
  \colhead{Parameter} & 
  \colhead{Credible Interval} & 
  \colhead{Units}
}
\startdata
\sidehead{\bf{Orbital Parameters}}
Orbital Period $P$ & $510.855\pm 0.014$ & days \\
Time of Inferior Conjunction $T\rm{conj}$ & $2452014.2\pm 0.13$ & JD \\
Time of Periastron $T\rm{peri}$ & $2452014.19\pm 0.16$ & JD \\
Eccentricity $e$ & $0.7008\pm 0.0018$ &  \\
Argument of Periapsis $\omega$ & $1.5696\pm 0.0056$ & radians \\
Velocity Semi-Amplitude $K$ & $311\pm 1$ & m s$^{-1}$ \\
\hline
\sidehead{\bf{Other Parameters}}
Mean Center-of-Mass Velocity $\gamma_{\rm apf}$ & $\equiv226.3573$ & m s$^{-1}$ \\
Mean Center-of-Mass Velocity $\gamma_{\rm CAT_{HES}}$ & $\equiv74.8982$ & m s$^{-1}$ \\
Linear Acceleration $\dot{\gamma}$ & $-0.05\pm 0.0017$ & m s$^{-1}$ d$^{-1}$ \\
Curvature $\ddot{\gamma}$ & $4.14e-06\pm 2.5e-07$ & m s$^{-1}$ d$^{-2}$ \\
Jitter $\sigma_{\rm apf}$ & $10.67^{+0.44}_{-0.42}$ & $\rm m\ s^{-1}$ \\
Jitter $\sigma_{\rm CAT_{HES}}$ & $13.85^{+0.92}_{-0.84}$ & $\rm m\ s^{-1}$ \\
\hline
\sidehead{\bf{Derived Posteriors}}
Mass $M_p\sin i$ & $11.67^{+0.45}_{-0.46}$ & M$_{\rm Jup}$ \\
Semi-Major Axis $a$ & $1.448^{+0.028}_{-0.029}$ & AU \\
\enddata
\end{deluxetable*}

\begin{figure*}
\centering 
  \includegraphics[width=0.7\textwidth]{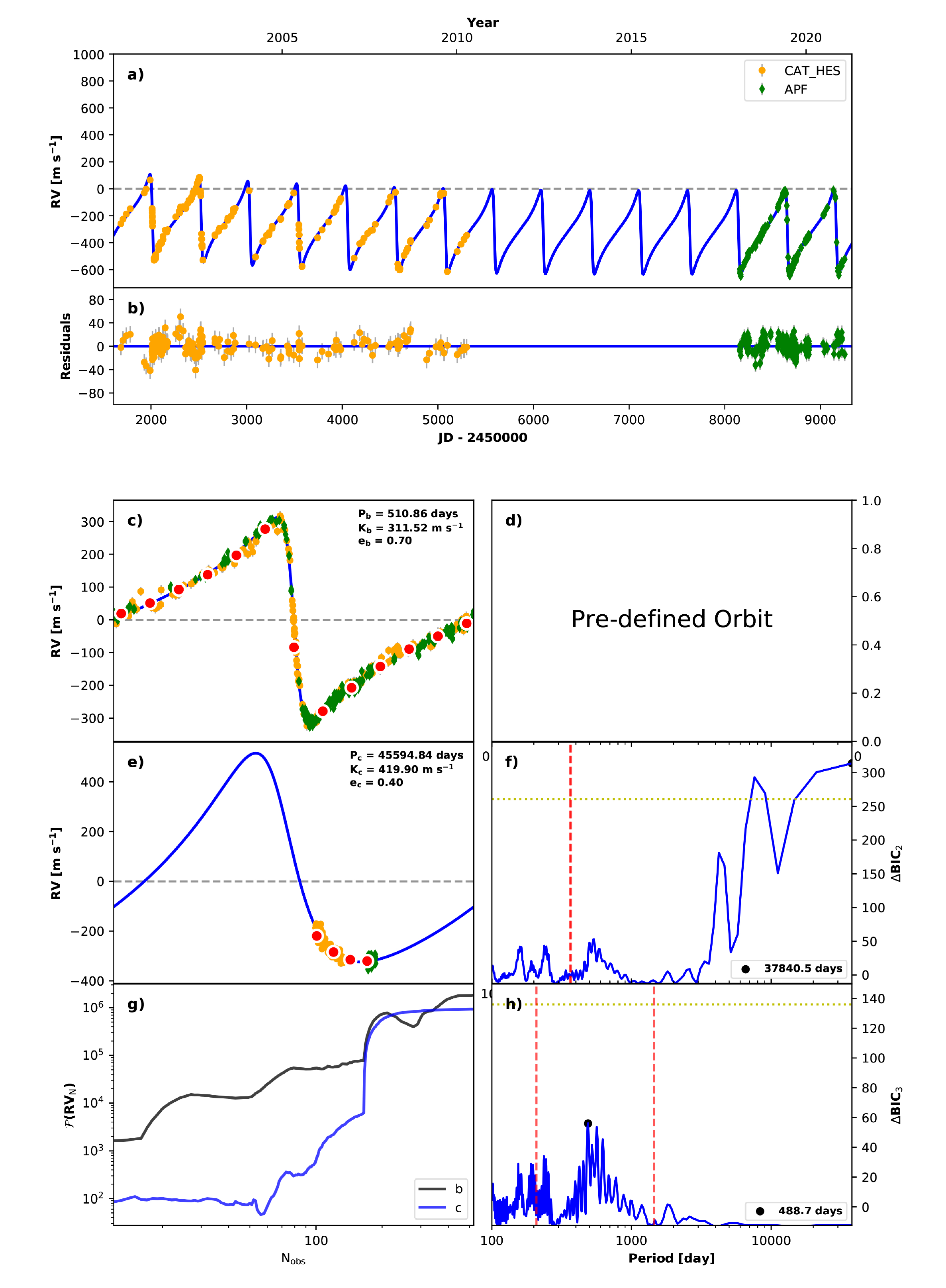}
\caption{ \texttt{RVSearch} results. Panels a, b, c show the known fit from Figure \ref{fig:137759}. Panel e shows the best fit of the new planet signal orbiting $\iota$~Dra with an estimated period of 45594.84~days. Panel f shows the periodogram with a signal at $\sim$37840.5~days. The horizontal dotted line indicates a false-alarm probability
(FAP) threshold of 0.001 (0.1\%). The vertical red dashed
line shows the location of a common alias caused by the
Earth’s orbital (annual) motions. Note the wide peak indicates a large uncertainty on the period of this signal. More data is needed to refine the orbital parameters of this additional planet. Panel g and Panel h show the running periodogram and the results from the residuals. The highest peak at 488.7 days does not exceed the FAP threshold of 0.001 and so it is concluded that there are
no further significant periodic signals present in the data. }
 \label{fig:137759RVSearch}
\end{figure*}

Using the iterative periodogram algorithm \texttt{RVSearch}
(Rosenthal et al. in prep), we searched for the period of the companion. \texttt{RVSearch} works by first defining the orbital frequency/period grid over which to search, with sampling such that the difference in frequency between adjacent grid points is $\frac{1}{2\pi \tau}$, where $\tau$ is the observational baseline. Using this grid, a goodness-of-fit periodogram was computed by fitting a sinusoid with a fixed period to the data for each period in the grid. The goodness-of-fit was measured as the change in the Bayesian Information Criterion (BIC) at each grid point between the best-fit 1-planet model with the given fixed period, and the BIC value of the 0-planet fit to the data. A power law was then fit to the noise histogram (50-95 percent) of the data and accordingly a BIC detection threshold corresponding to an empirical false-alarm probability of 0.0003 was extrapolated. If one planet was detected, a final fit to the one-planet model with all parameters free was completed, and the BIC of that best-fit model recorded. Then a second planet was added to the RV model and another grid search conducted, leaving the parameters of the first planet free to converge to a more optimal solution. In this case the goodness-of-fit was computed as the difference between the BIC of the best-fit one-planet model, and the BIC of the two-planet model at each fixed period in the grid. The detection threshold was set in the manner described above and this iterative search continued until the n+1th search ruled out additional signals. 
For $\iota$~Dra one significant companion signal was detected by the algorithm. The periodogram
resulting from this analysis is shown in Figure \ref{fig:137759RVSearch} panel f. The horizontal dotted line indicates a false-alarm probability
(FAP) threshold of 0.001 (0.1\%). The vertical red dashed
line shows the location of a common alias caused by the
Earth’s orbital (annual) motions. Panel e shows the best fit of the signal with an estimated period of $\sim$~45594~days, eccentricity of $\sim$~0.4, semi-amplitude of $\sim$~420~m~s$^{-1}$ and $M_p\sin i$~of~$\sim$~38~M$_{Jup}$. This detection is beyond the baseline of the RV data and so there is a large uncertainty associated with this period. To refine the parameter space of the companion we run both a dynamical analysis with \texttt{MEGNO} and then further constrain the orbits with \texttt{The Joker}.

\subsection{Dynamical Analysis With \texttt{MEGNO}}
\label{dynamics}

To constrain the parameter space of the additional orbiting body, we performed a dynamical simulation using the \texttt{MEGNO} (Mean Exponential Growth of Nearby Orbits) chaos indicator \citep{cincotta2000} to determine the range of semi-major axis and eccentricity configurations that this second body could have. The \texttt{MEGNO} simulation was carried out within the N-body package \texttt{REBOUND} \citep{rein2012a} with the symplectic integrator \texttt{WHFast} \citep{Rein2015}. For planetary systems, the \texttt{MEGNO} indicator is useful in distinguishing the quasi-periodic or chaotic orbital time evolution of planetary bodies within the system \citep{hinse2010}, where a chaotic state for a planet is less likely to maintain long-term orbital stability. For our simulation, we explored the possible orbital configurations of this potential outer companion by varying its semi-major axis and eccentricity value. The range of semi-major axis was tested between 8 and 40 AU and for eccentricity, we tested a range from 0 and 0.75. The lower limit of the semi-major axis is provided by the baseline of observations, as a full orbit has not been completed. The upper limit of eccentricity was provided by the initial \texttt{JOKER} fit (see Section~\ref{Joker}). The mass of the outer orbiting body was assumed to be 38 Jupiter masses (M$_{jup}$). 
All bodies were assumed to be co-planar with edge-on inclination and the outer companion was assigned an argument of periastron value of 326 degrees derived from the \texttt{JOKER} fit. The \texttt{MEGNO} simulation was run for each grid point for 20 million years integration time with a time step of 0.035 years ($\sim$13 days). The time step is equivalent to $1/40$ the orbital period of planet b, and is half of the recommended time step \citep{duncan1998} to ensure enough sampling when the highly eccentric planet b passes through periastron.

\begin{figure}[!t]
\centering
\includegraphics[width=\linewidth]{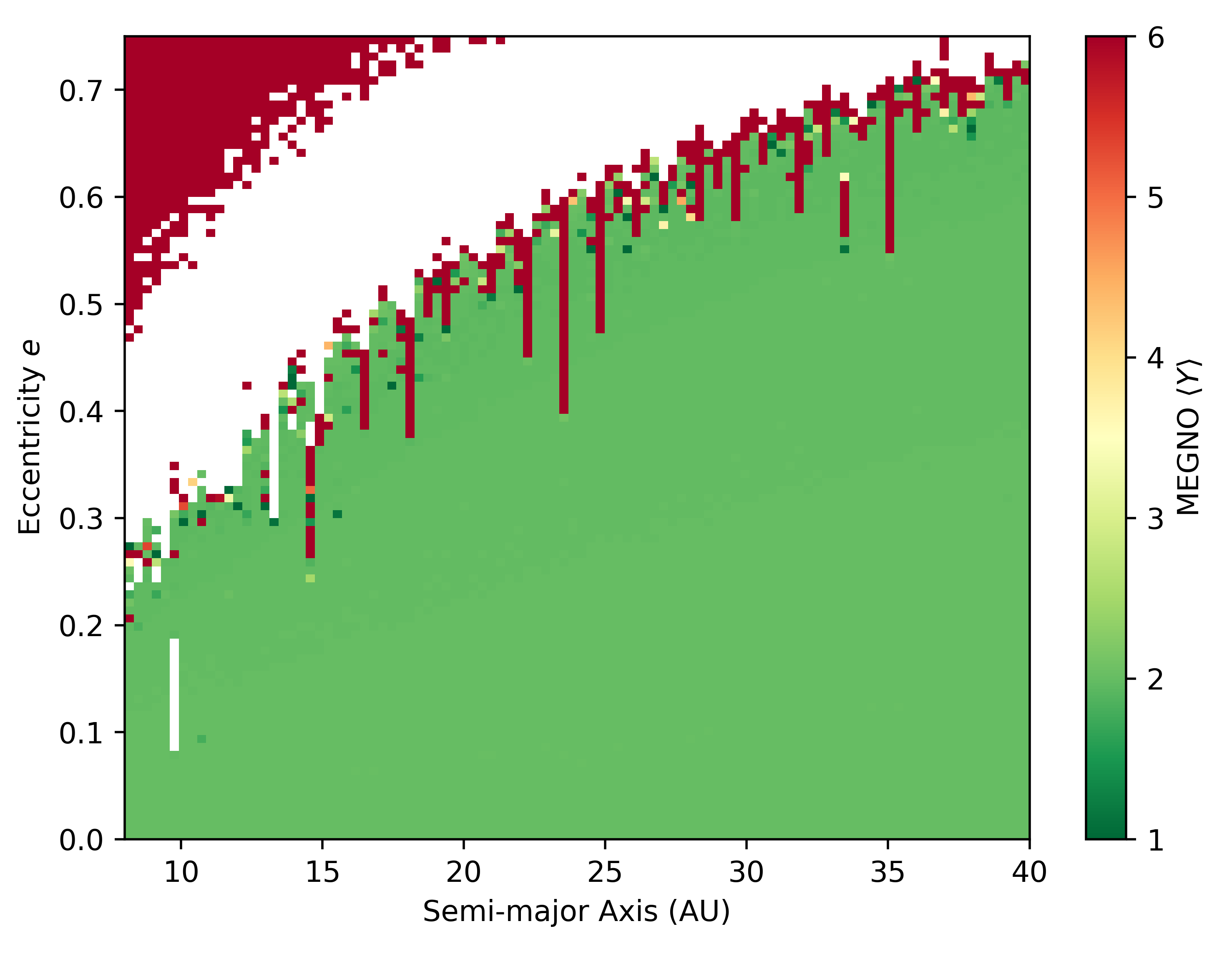}
\caption{\texttt{MEGNO} simulation result showing the possible orbital configurations of the potential outer companion. Grids in green with value around 2 indicate non-chaotic results and are dynamically viable locations. Grids in other colors represent chaos and are not favorable locations for the outer companion. Areas in white indicate NaN \texttt{MEGNO} values where early termination was caused by irregular events such as close encounters and/or collisions. \label{fig:MEGNO}}
\end{figure}

Shown in Figure \ref{fig:MEGNO} is the result of the simulation. The horizontal and vertical axes represent the range of semi-major axis and eccentricity that we tested. Each grid is color-coded based on the final \texttt{MEGNO} value for that specific configuration, where a \texttt{MEGNO} value around 2 (green) indicates non-chaotic results and planets all undergo quasi-periodic motion. Higher \texttt{MEGNO} values represented by warmer colors indicate chaos results for the system, and early termination with NaN \texttt{MEGNO} values caused by irregular events such as close encounters and collisions are marked in white. Locations with \texttt{MEGNO} values far from 2 are not favorable locations for the potential outer companion. The simulation indicates that the system would be unlikely to be in a chaotic state if the outer companion orbits close to the lower limit of semi-major axis with a low eccentricity. But other configurations with higher eccentricities become available at larger orbital separations, except several locations indicated by the white or red vertical bars where resonances may exist.

\subsection{Further Constraints On Outer Companion}
\label{Joker}

 To further refine the parameter space, we used \texttt{The Joker} \citep{Price_Whelan_2017} to predict orbital solutions for the additional body. 
 \texttt{The Joker} is a Monte Carlo sampler that employs von Neumann rejection sampling to model RV variations for two-body systems \citep{Price_Whelan_2017}. Our interest in constraining the orbital properties of the outer companion required us to first remove the signal of the inner planet from the RV observations. We subtracted the maximum-likelihood fit from the \texttt{Radvel} analysis (Section~\ref{RV_analysis}) but did not include the fitted values of acceleration (linear trend and/or curvature). This left a time series of RVs that contained only the trend from the outer companion.
 
 In fitting these RV data with \texttt{The Joker}, we applied the default priors. Companion orbital period was assumed to be log normal between 7500 days (roughly the baseline of observations) and 100000 days. The prior over companion eccentricity was a Beta distribution with shape parameters $\alpha = 0.867$ and $\beta = 3.03$, which describes the known exoplanet samples at long orbital periods \citep{kipping2013b}. The argument and phase of periastron as well as the semi-amplitude and systemic velocity all had uniform, non-informative priors. Lastly, we fixed the RV jitter to the value derived in the \texttt{Radvel} analysis (Section~\ref{RV_analysis}).
 
\begin{figure}
\centering 
\subfloat{%
  \includegraphics[width=1.0\columnwidth]{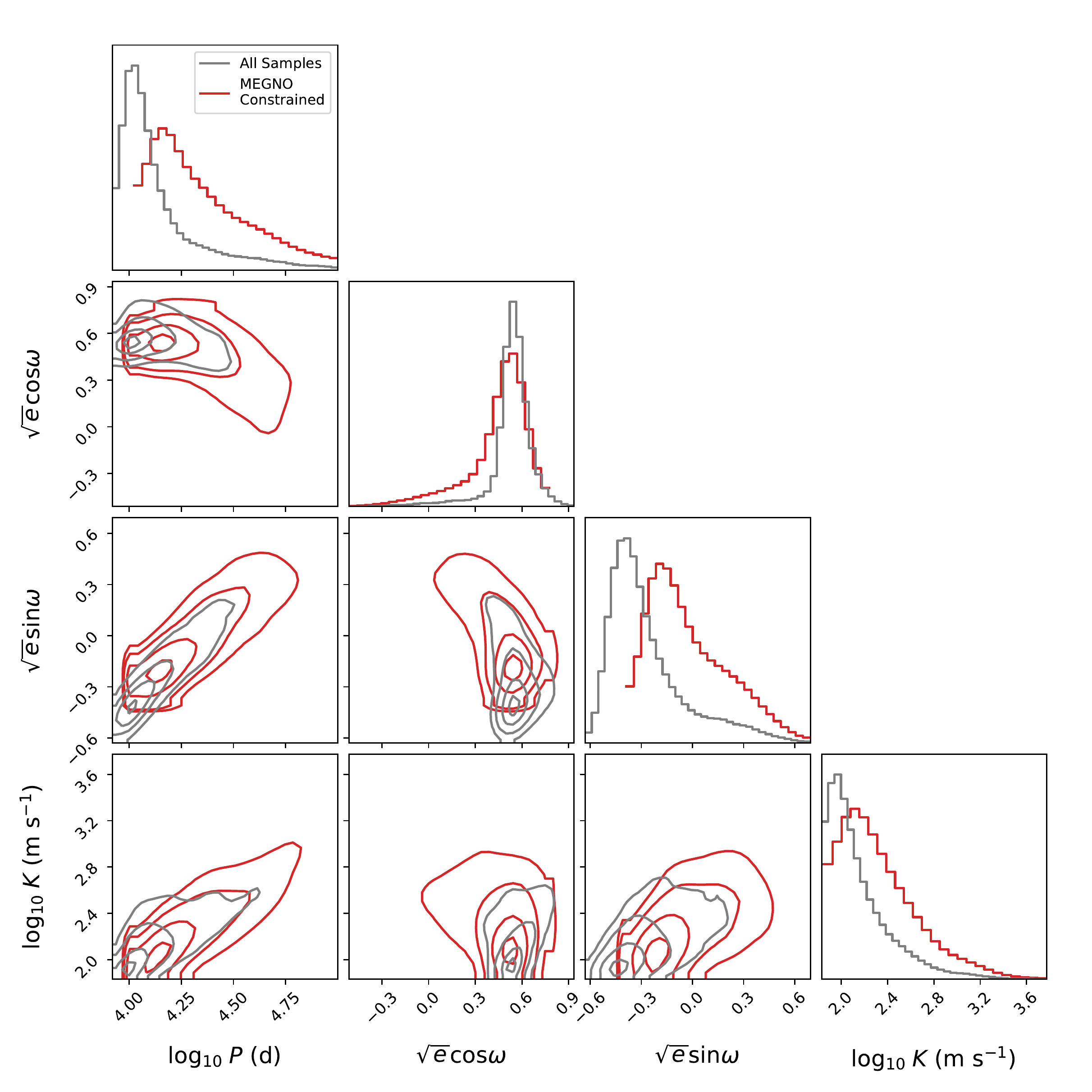}%
}\qquad
\subfloat{%
  \includegraphics[width=0.8\columnwidth]{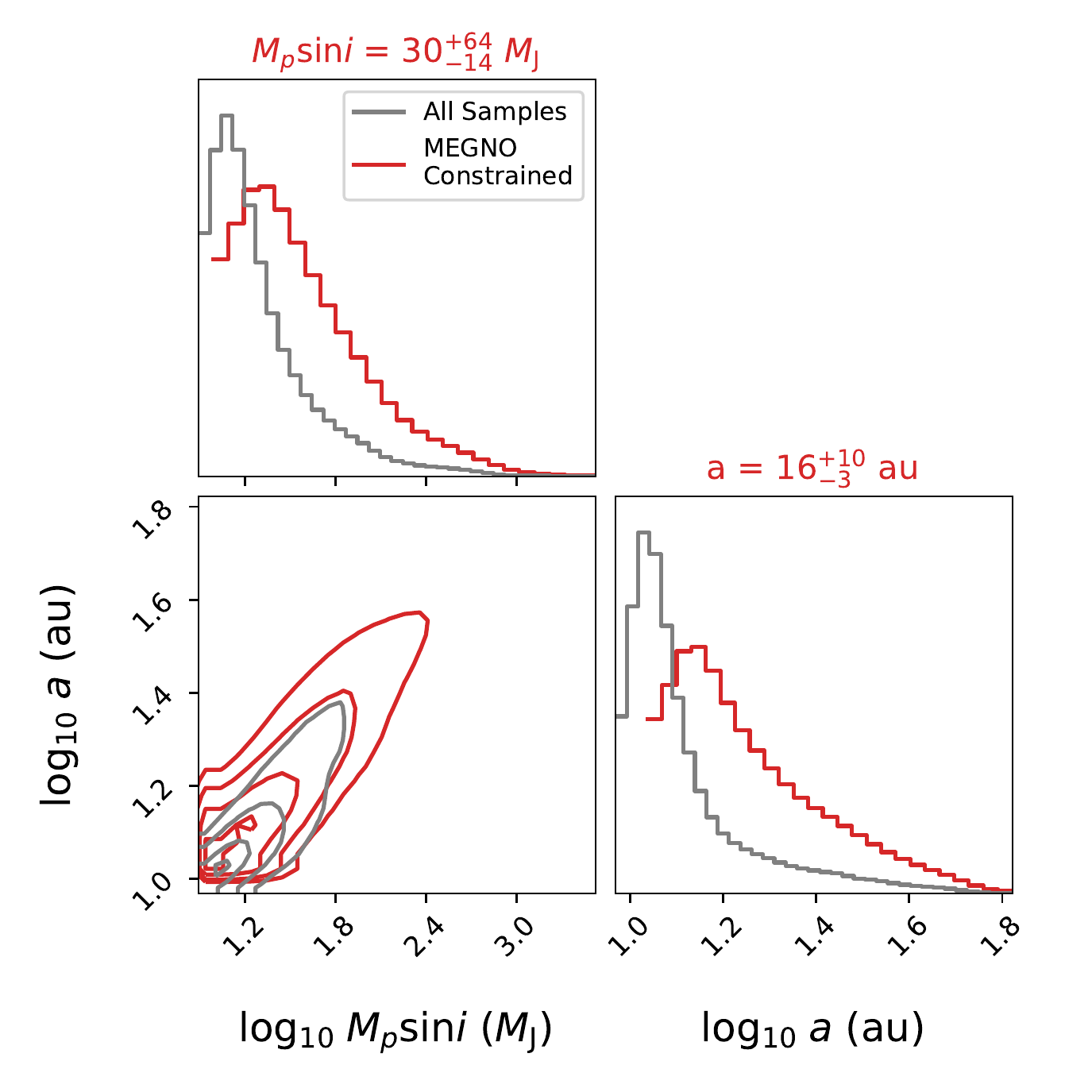}%
}
\caption{Top: \texttt{The Joker} fitted posteriors and Bottom: derived posteriors for mass ($M_p\sin i$) and semi-major~axis~($a$). Gray posteriors are results from \texttt{The Joker} without constraints from the \texttt{MEGNO} analysis. Red posteriors include constraints from the \texttt{MEGNO} analysis. 
The inclusion of the \texttt{MEGNO} constraints resulted in the removal many of the shorter-period, highly eccentric solutions. The orbital period, semi-major axis, minimum companion mass, and semi-amplitude were all pushed toward higher values. In particular, the minimum mass moved from the planetary mass regime and into the brown dwarf regime.
\label{fig:Joker_post}}
\end{figure}

 Using \texttt{The Joker}, we made $2^{30}$ ($\sim 1.1\times10^9$) draws from the prior distributions, of which 8912 survived.  
 We show the posteriors comprised of the surviving samples in Figure~\ref{fig:Joker_post}. The posterior for companion orbital period peaks near 10000 days but has a long, low probability tail out to longer values. We found that this tail continued out to whatever maximum period value we chose, which was not surprising given that small fraction of the orbit our data cover. The shorter-period solutions generally required higher eccentricity whereas longer-period solutions defaulted to the prior and were more circular. Companion minimum mass peaked around 11~$M_{\rm J}$ but also had a long, low-probability tail to higher values. 
 
 So far, our analysis has not considered the constraints derived by the \texttt{MEGNO} analysis. However, as shown in Figure~\ref{fig:MEGNO}, this simulation provides a constraint in eccentricity-semi-major axis space. There is an \emph{envelope} of low \texttt{MEGNO} $<Y>$ values that favor lower eccentricity, larger orbits. We approximated this envelope as the interface between the white and red/green regions in Figure \ref{fig:MEGNO} and used it to divide the posteriors from \texttt{The Joker}. This kind of analysis has been employed previously to interpret results from \texttt{The Joker} in combination with additional limiting information \citep{dalba2020b}. The resulting posteriors are overplotted on those for all surviving draws in Figure~\ref{fig:Joker_post}. The primary effect of including the chaos indicator results is to remove many of the shorter-period, highly eccentric solutions. This pushes the orbital period, semi-major axis, minimum companion mass, and semi-amplitude all toward higher values. Specifically, the minimum companion mass moves out of the planetary mass regime and into the brown dwarf (or substellar) regime. 
 
\begin{figure}[!ht]
\centering
\includegraphics[width=\linewidth]{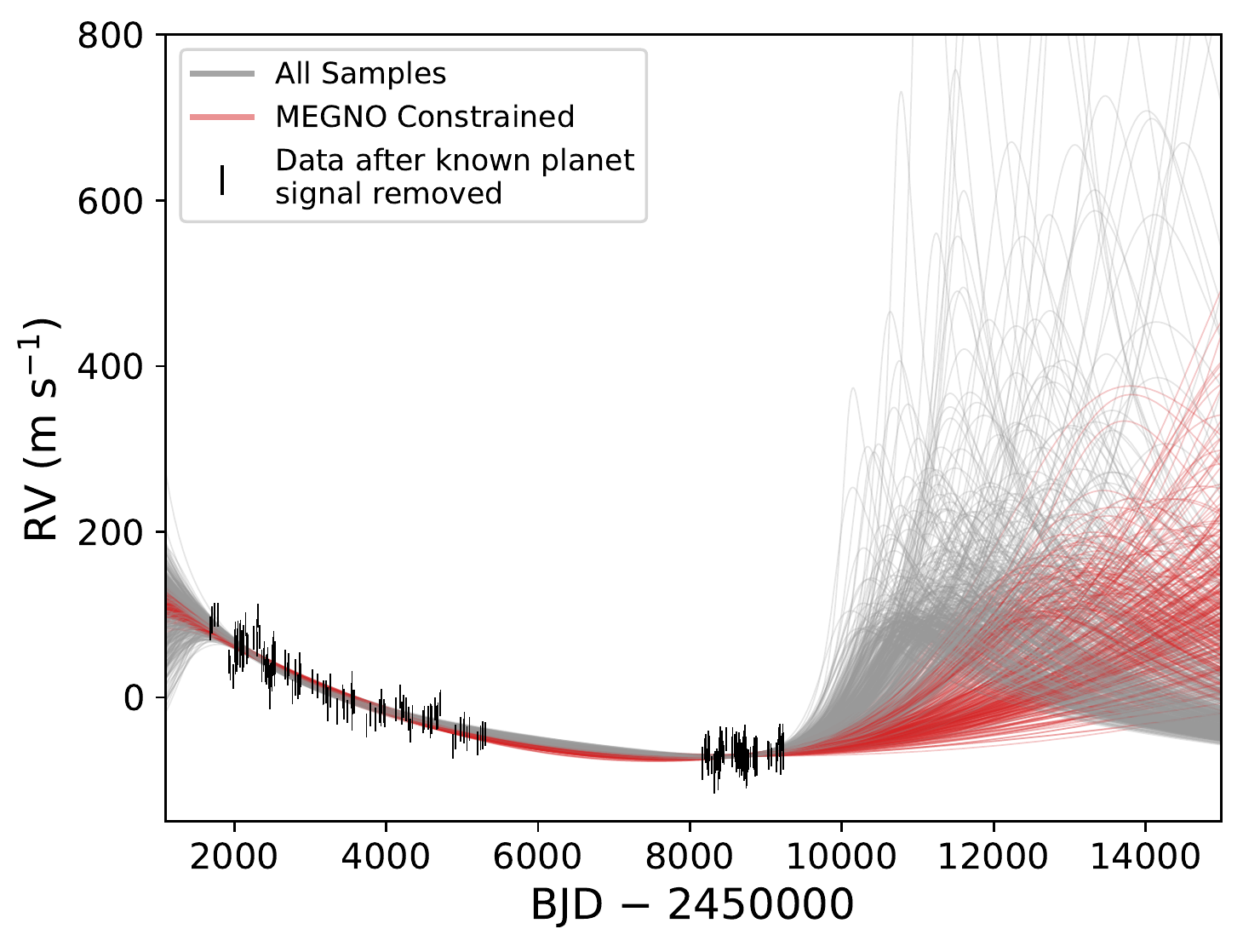}
\caption{A representative subset of the surviving \texttt{The Joker} orbits. The orbits are a draw of $\sim$10\% of the full set of surviving draws. Black vertical lines represent the time series RV data after the subtraction of the known planet signal. The gray curves represent orbital solutions which do not consider the \texttt{MEGNO} constraints, red curves are those that include the \texttt{MEGNO} constraints. Inclusion of the \texttt{MEGNO} analysis pushes the orbits to longer period, less eccentric solutions. \label{fig:Joker_orbits}}
\end{figure}

 In Figure~\ref{fig:Joker_orbits}, we display a representative subset of the orbits corresponding to the surviving posterior draws. We also show the time series RV observations after the subtraction of the known planet signal. The gray curves, which do not consider the \texttt{MEGNO} constraints are noticeably more eccentric and shorter-period than the red curves, which are consistent with the \texttt{MEGNO} analysis.

\setlength{\belowcaptionskip}{-8pt}

\bigskip


\subsection{Combined Radial Velocity and Astrometry Analysis}
\label{Astrometry}

By combining absolute astrometry with RV, the true motion of the star can be observed and planet mass and orbital parameters further constrained. RV measurements probe the line-of-sight acceleration while absolute astrometry measures the orthogonal component. For $\iota$~Dra~c, the RV curvature sets a lower limit on the mass, while the lack of a significant acceleration in the absolute astrometry provides an upper limit. We used \texttt{orvara}, an open-source Python package created by \citet{orvara2021} (details found therein) that performs comprehensive joint MCMC analysis to determine orbital fits for planetary (or binary star)
systems using a combination of Hipparcos \citep{ESA1997, van_Leeuwen_2007} and Gaia \citep{prusti2016, GaiaDR2, GaiaDR3_2021} astrometry, RV, and/or
relative astrometry. 

\begin{figure}[!ht]
\centering
\includegraphics[width=\linewidth]{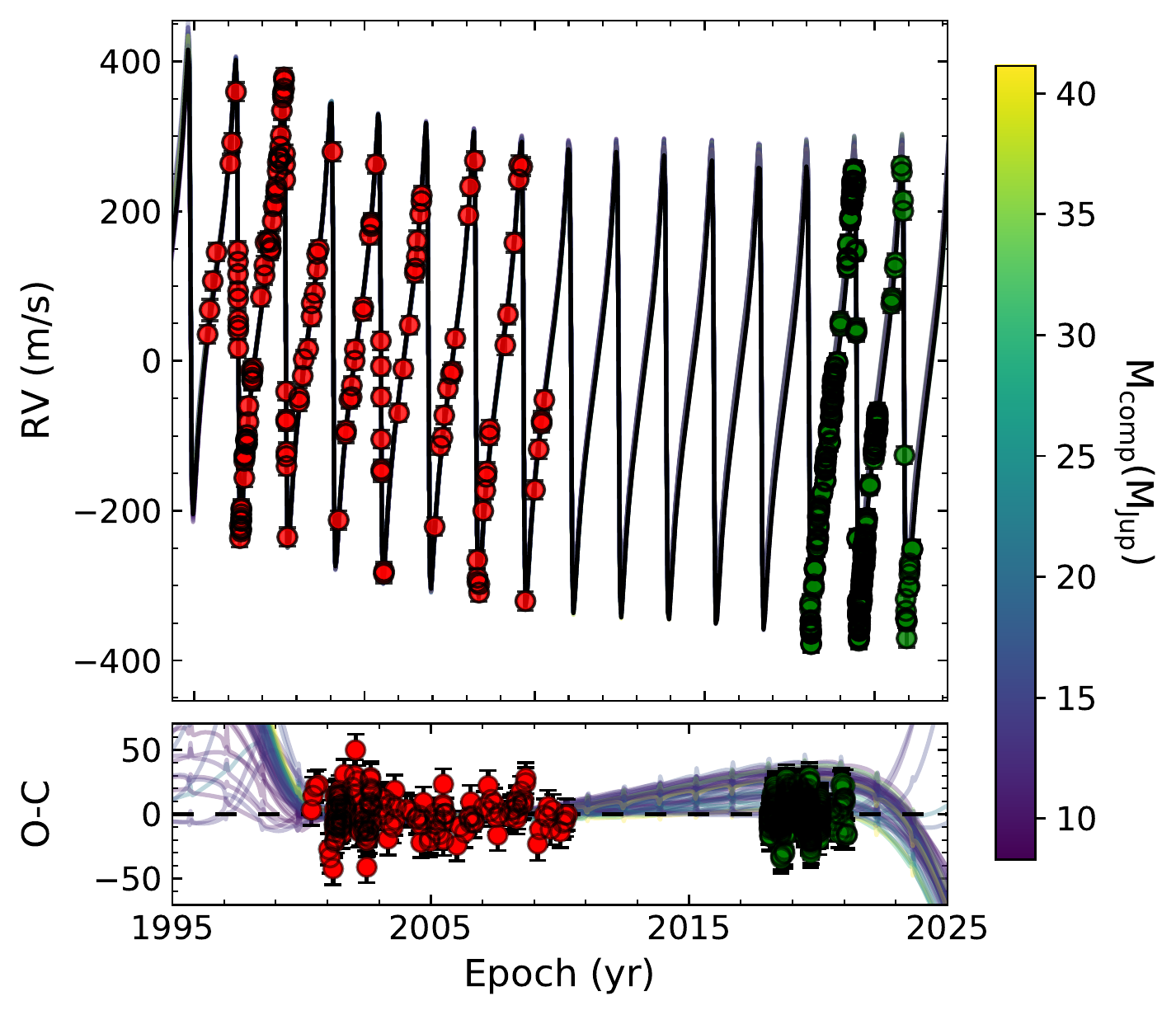}
\caption{\texttt{orvara} fit of the $\iota$~Dra system. Data points in red are those from the Coud\'{e} Auxiliary Telescope (CAT) with the Hamilton \'{E}chelle Spectrograph (HES); those in green were taken using the Automated Planet Finder (APF) with the Levy spectrometer. The thick black line shows the shows the highest likelihood orbital solution. The fit produced by \texttt{orvara} agrees with the \texttt{RadVel} fit shown in Figure \ref{fig:137759}. The bottom panel shows the observed-calculated (O-C) residuals, along with colored lines indicating 50 random orbits from the posterior probability distribution. The colorbar on the right of the plot indicates the mass of the outer companion ($M_{\rm comp}$) that each of these colors represent.   \label{fig:RV_orvara}}
\end{figure}

The RV data along with absolute astrometry from Gaia and Hipparcos was fit by \texttt{orvara} and the best fit solution is shown in Figure \ref{fig:RV_orvara}.
The top panel shows the RV observations from the Lick Observatory CAT with the HES in red and the APF in green, with the thick black line showing the highest likelihood orbital solution. The bottom panel represents the corresponding
Observed-Calculated (O-C) residuals. The O-C residuals indicate the
deviation of the observed value from the most-likely orbit. The bottom panel also includes 50 colored lines. These represent 50 orbits chosen randomly from  the posterior probability distribution, the colors corresponding to the mass of the companion (M$_{\rm comp}$) as indicated by the colorbar on the right.
The preferred fit by \texttt{orvara} agrees with the \texttt{RadVel} solution of $\iota$~Dra~b in Section \ref{RV_analysis}, including the $M_p\sin i$ value of 11.82$^{+0.42}_{-0.41}$ which agrees with the \texttt{RadVel} estimate in Table \ref{tab:Radvel2planet}. As both Hipparcos and Gaia fits to the proper motion of stars are integrated over each instrument's mission baseline ($\sim3.5$ years for Hipparcos, and 34 months for Gaia EDR3), rather than measuring instantaneous proper motions, astrometry with these instruments is not sensitive to planets with periods shorter than this baseline, like $\iota$~Dra~b. For this reason it is expected that the result from \texttt{orvara} for the inner planet would agree with the \texttt{RadVel} solution. Astrometry is, however, extremely useful when fitting a long-period object like the outer companion.

\begin{figure*}
\centering 
\subfloat{%
  \includegraphics[width=1.5\columnwidth]{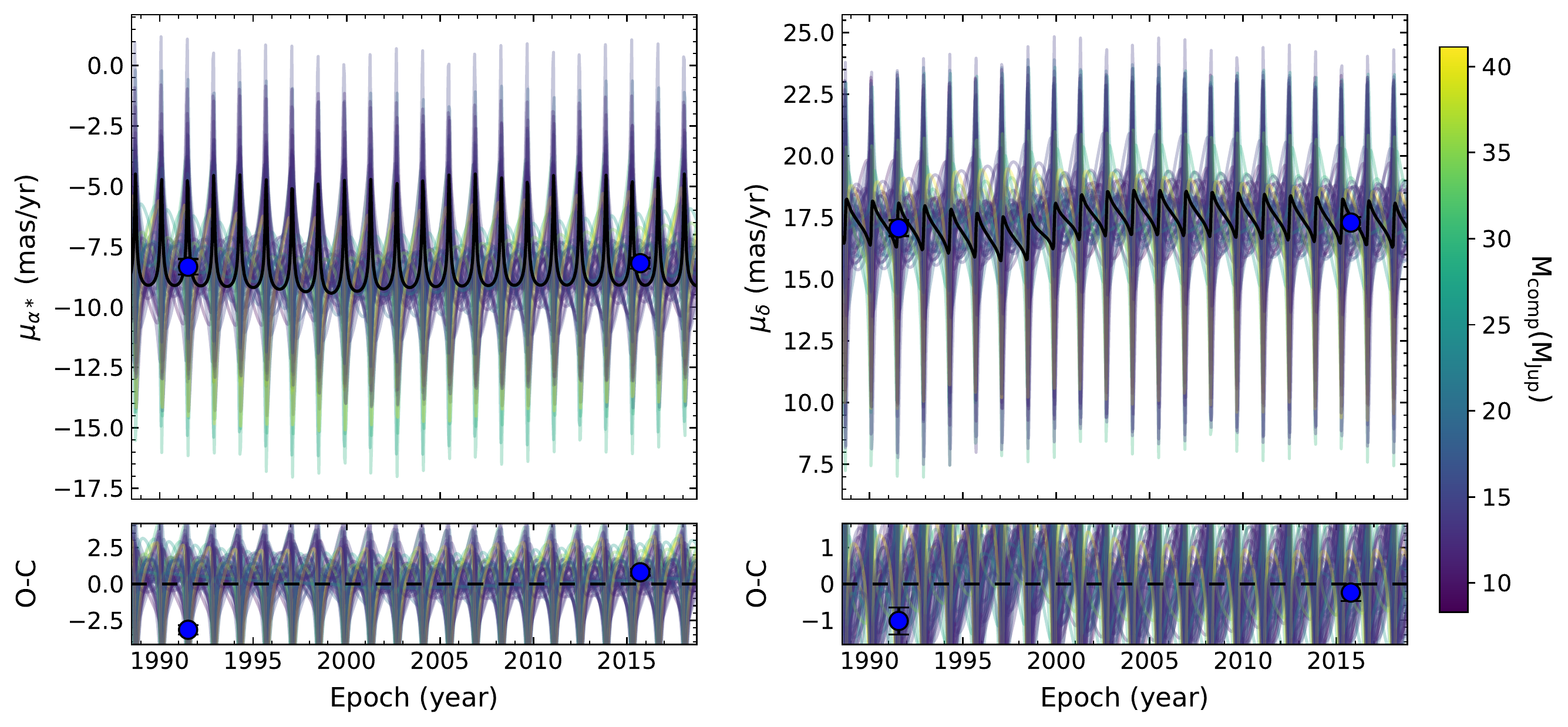}%
}\qquad
\subfloat{%
  \includegraphics[width=1.5\columnwidth]{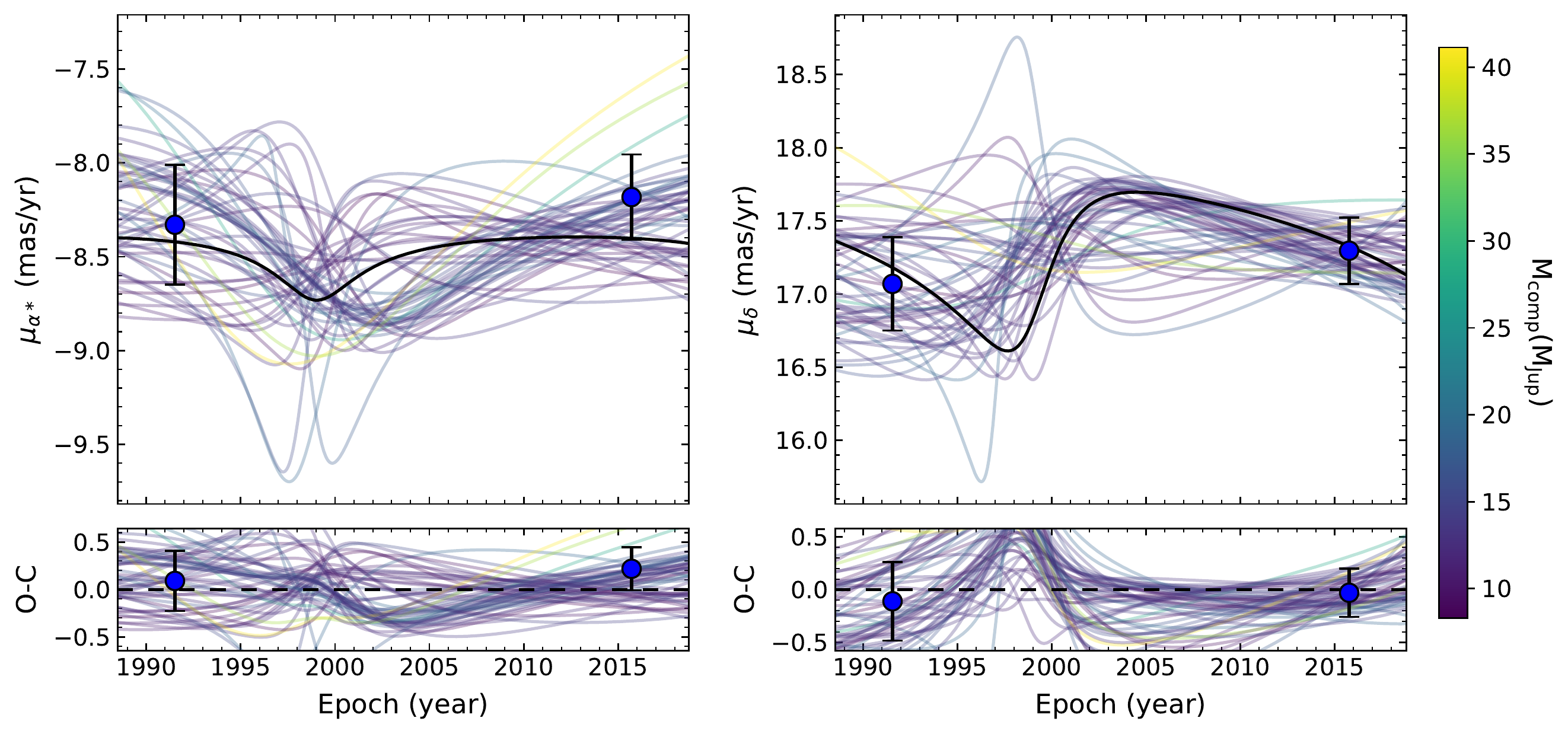}%
}
\caption{ \texttt{orvara} proper motion of $\iota$~Dra due to both companions (top) and due to just the outer companion (bottom), in right ascension ($\mu_{a^*}$) on the left and declination ($\mu_{\delta}$) on the right. The black line represents the best fit orbit in the MCMC chain whereas the other lines represent 50 random draws. The color bar on the right indicates the outer companion mass ($M_{\rm comp}$) for each orbital draw. The lower panels of each run represent the Observed-Calculated (O-C) residual. 
\label{fig:proper_motion}}
\end{figure*}

\texttt{orvara} uses the Hundred
Thousand Orbit Fitter (\texttt{htof}, Brandt et al. submitted) to fit the proper motion of a star. \texttt{htof} computes synthetic
Hipparcos and Gaia catalog positions and proper
motions and then compares this to the absolute astrometry from the Hipparcos-Gaia Catalog
of Accelerations (HGCA) \citep{Brandt_2018,Brandt_2021}. We used the EDR3 version of the HGCA for $\iota$~Dra.  The proper motion of the $\iota$~Dra system in right ascension ($\mu_{a^*}$) and declination ($\mu_{\delta}$) due to both companions, and also from just the outer companion,  are in Figure \ref{fig:proper_motion}. Again, the black line represents the best fit orbit in the MCMC chain, whereas the other colored lines represent 50 random draws with masses corresponding to the colorbars on the right. Due to the integration period of both Gaia and Hipparcos exceeding the period of $\iota$~Dra~b, little information is gained by including proper motion analysis for this planet, as can be seen in the top panels of Figure \ref{fig:proper_motion}. The bottom panels, showing the proper motion of $\iota$~Dra due to the outer companion, are much more informative and show the orbital constraints determined from the inclusion of proper motion for this object clearly. Note: The mean proper motion was used to compute the MCMC chain, but
is not shown in the proper motion plots.  It is a constraint on the integrated proper motion between $\approx$1991 and $\approx$2016.

The result of our comprehensive joint
MCMC analysis of the RV and astrometric data is presented in the corner plot of Figure \ref{fig:prop_c_orvara} and Table \ref{tab:orvara}.  This includes derived posterior probabilities for the outer companion's semi-major axis $a$, eccentricity $e$, and inclination $i$, mass M$_{sec}$, and the mass of the
primary star M$_{\rm pri}$. Rather than using the default priors of 1/M$_{\rm sec}$ and 1/$a$ for the companion mass and semi-major axis respectively, we use uniform priors for each to avoid causing a bias to low-mass and low semi-major axis solutions.
The preferred solution from \texttt{orvara} gives the outer companion a mass of ${17.0}_{-5.4}^{+13}$ M$_{\rm Jup}$. This  puts the outer companion on the border of the planet and brown dwarf regimes. This mass estimate agrees with the predicted mass from \texttt{The Joker} analysis in Subsection \ref{Joker} and confirms that the outer companion is a sub-stellar object.

\setlength{\belowcaptionskip}{2pt}

\begin{deluxetable*}{lrrr}[ht]
\tablecaption{\label{tab:orvara} \texttt{orvara} Posteriors}
\tablehead{
  \colhead{Parameter} & 
  \colhead{Planet b} & 
  \colhead{Planet c} & 
  \colhead{Units}
}
\startdata
\sidehead{\bf{Fitted Parameters}}
Companion Mass ($M_{\rm sec}$) & ${16.4}_{-4.0}^{+9.3}$ & ${17.0}_{-5.4}^{+13}$ & M$_{\rm Jup}$ \\
Semi-major Axis ($a$) & ${1.453}_{-0.026}^{+0.026}$ & ${19.4}_{-7.7}^{+10}$ & AU \\
$\sqrt{e}\sin\omega$ & ${0.8373}_{-0.0010}^{+0.0010}$ & ${0.44}_{-0.64}^{+0.24}$ &  \\
$\sqrt{e}\cos\omega$ & ${0.0015}_{-0.0043}^{+0.0044}$ & ${0.46}_{-0.23}^{+0.15}$ &  \\
Inclination & ${46}_{-19}^{+27}$ & ${86}_{-19}^{+19}$ & deg \\
Ascending node & ${87}_{-60}^{+64}$ & ${107}_{-59}^{+44}$ & deg \\
Mean longitude & ${173.18}_{-0.24}^{+0.23}$ & ${108.7}_{-13}^{+9.4}$ & deg \\
\hline
\sidehead{\bf{Derived Parameters}}
Period & ${1.398643}_{-0.000035}^{+0.000035}$ & ${68}_{-36}^{+60}$ & yrs \\
Argument of Periastron & ${89.90}_{-0.30}^{+0.30}$ & ${62}_{-32}^{+262}$ & deg \\
Eccentricity & ${0.7010}_{-0.0017}^{+0.0016}$ & ${0.455}_{-0.084}^{+0.12}$ &  \\
Semi-major Axis & ${47.26}_{-0.83}^{+0.83}$ & ${630}_{-250}^{+328}$ & mas \\
$T_0$ & ${2455590.17}_{-0.13}^{+0.13}$ & ${2476000}_{-13000}^{+22000}$ & JD\\
Mass ratio & ${0.0100}_{-0.0024}^{+0.0058}$ & ${0.0105}_{-0.0034}^{+0.0080}$ & \\
$M_p\sin i$ & ${11.82}_{-0.41}^{+0.42}$ & ${15.6}_{-5.1}^{+14}$ & M$_{\rm Jup}$ \\
\hline
\sidehead{\bf{Other Parameters}}
Jitter & ${11.42}_{-0.33}^{+0.36}$ &  & m s$^{-1}$ \\
Stellar Mass ($M_{\rm pri}$) & ${1.551}_{-0.078}^{+0.083}$ &  & M$_{\rm sun}$ \\
Parallax & ${32.5224}_{-0.0016}^{+0.0010}$ & & mas \\
Barycenter Proper Motion RA & ${-8.23}_{-0.25}^{+0.60}$ & & mas\,yr$^{-1}$ \\
Barycenter Proper Motion DEC & ${17.22}_{-0.33}^{+0.16}$ & & mas\,yr$^{-1}$ \\
RV Zero Point CAT\_HES & ${-14}_{-48}^{+29}$ & & m\,s$^{-1}$ \\
RV Zero Point APF & ${-143}_{-49}^{+21}$ & & m\,s$^{-1}$ \\
\enddata
\end{deluxetable*}

\begin{figure*}[!ht]
\setlength{\belowcaptionskip}{2pt}
\centering
\includegraphics[width=1.3\columnwidth]{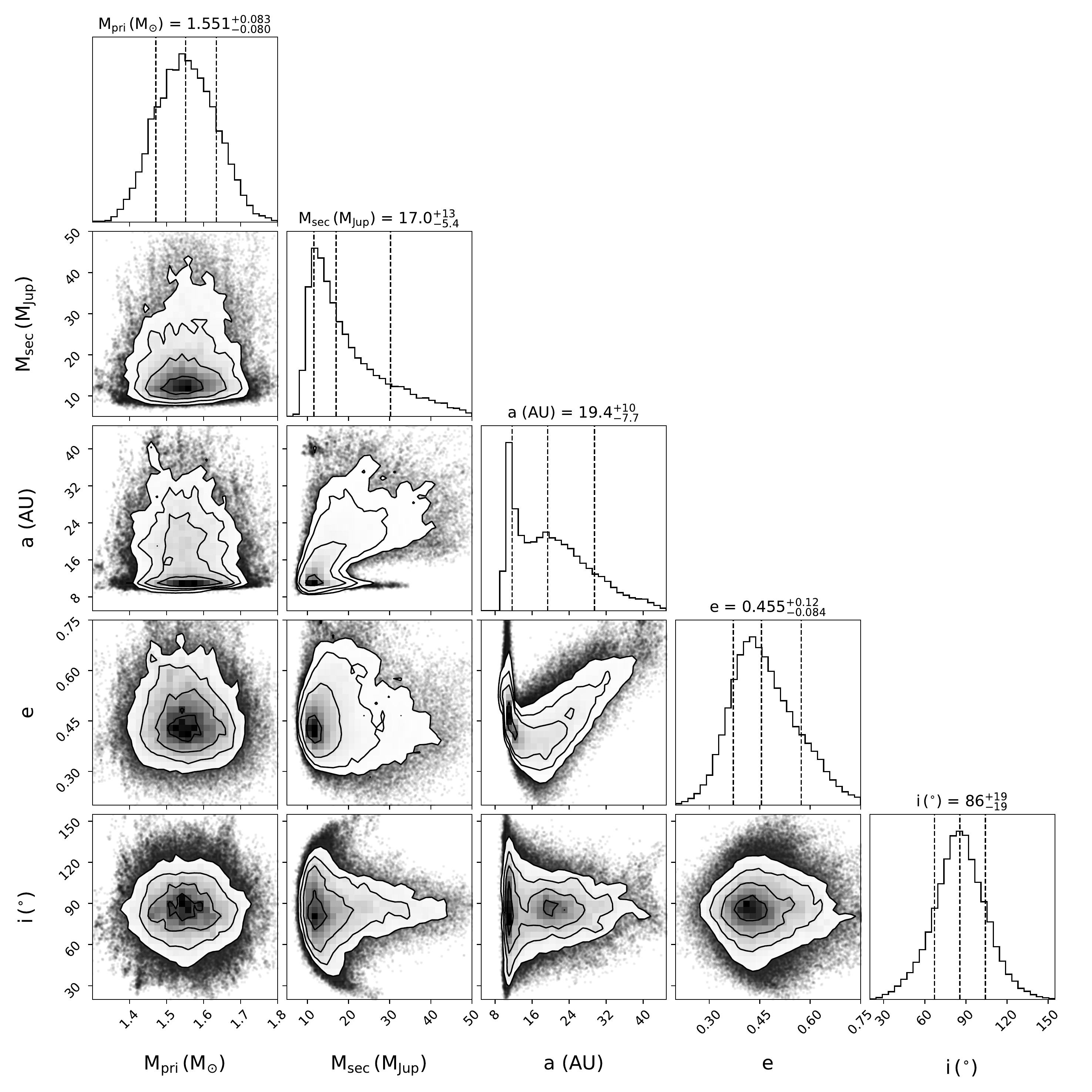}
\caption{\texttt{orvara} posteriors for the mass of the star ($M_{\rm pri}$), mass of the outer companion ($M_{\rm sec}$), semi-major axis ($a$), eccentricity ($e$) and inclination ($i$). Note the prior for $M_{\rm pri}$ was taken from our results in Table \ref{tab:starprop}.  By combining RV and proper motion analysis \texttt{orvara} provides a further constraint on the outer companion, with a posterior mass of $17.0^{+13}_{-5.4}$\,$M_{\rm Jup}$, which puts the companion on the boundary between the planet and brown dwarf regimes. These results are tabulated in Table \ref{tab:orvara}. \label{fig:prop_c_orvara}}
\end{figure*}


\section{Discussion}
\label{discussion}

Stellar evolution can have an important impact on the dynamical evolution of the planets in a system \citep{Jones2014, Damiani2018, Grunblatt2019}, including planets that lie within a dramatically evolving Habitable Zone \citep{gallet2017a, Farr2018}. The discovery of planets in eccentric orbits around evolved stars is critically important for diagnosing the source of such eccentricities \citep{wittenmyer2017a, Bergmann2021, Grunblatt2018}, whether it be due to mass-loss \citep{Soker2001, adams2013} or planet-planet scattering
\citep{kane2014b,carrera2019b}. Thus, the prospect of $\iota$~Dra being a multi-planet system containing significantly eccentric orbits becomes a useful case study in the evolutionary history of planets orbiting evolved stars.

Precision RV surveys for exoplanets have now been operating for several decades, extending the period sensitivity toward Saturn analogs \citep{montet2014,wittenmyer2020b}. However, the detection and characterization of planetary orbits outside the current observing window remains challenging due to the difficulties in reliably extracting Keplerian orbital parameters from data with partial phase coverage \citep{dalba2020b}. A notable exception lies in the case of HR~5183b, whose eccentric $\sim$74 year orbit was observed during periastron passage \citep{blunt2019}, almost entering the Habitable Zone of the system \citep{kane2019e}. The orbital period of the additional companion to $\iota$~Dra described here, though relatively unconstrained, is similar in value to that of HR~5183b (see Table~\ref{tab:orvara}). Continued low-cadence monitoring of the system will constrain the eccentricity of the orbit which, in turn, will provide further improvements to the orbital period without the need to observe a complete orbit.

Jitter estimates from well sampled RV timeseries of giants are relatively rare. Studies of RV jitter made by \citet{Tayar2019} and \citet{Luhn2020} predict RV RMS for stars similar to $\iota$~Dra of $12-16 m s^{-1}$ and $15m s^{-1}$ respectively. The jitter terms in Table \ref{tab:Radvel2planet} are estimated to be slightly smaller than, but broadly consistent with the predictions from these studies, with jitter $\sigma_{\rm apf}$ $\sim 10 m s^{-1}$ and jitter $\sigma_{\rm CAT_{HES}}$ $\sim 13.5 m s^{-1}$. With a $\log g$ of $2.48\pm0.01$, $\iota$~Dra is on the boundary of the stellar samples included in these studies, future studies into the predictions of stellar jitter should be extended to include stars of lower $\log g$.

As noted by \citet{kane2010a}, the large stellar radius of $\iota$~Dra, combined with the favorable orientation of the highly eccentric planetary orbit, yield a transit probability of $\sim$16\% for the known inner planet. TESS observations of $\iota$~Dra during the initial mission cycles did not coincide with the expected time of inferior conjunction for $\iota$~Dra~b, and thus a transit of this planet has not yet been ruled out. However, subsequent TESS observations will revisit this star, providing further opportunities to detect a possible transit. The next window for this potential transit will occur at BJD 2459677.03$\pm$0.13 (April 7, 2022), when TESS will be in its 4+ year, and scheduled to observe Sector 50. Provided Sector 50 is observed on schedule, $\iota$~Dra will be observed during this transit window and so any transit of $\iota$~Dra~b should be observed. 

As a companion to a bright $V=3.3$ magnitude star, the brown dwarf orbiting $\iota$~Dra could be a prime candidate for future direct imaging missions. At 30.74~pcs away and with a semi-major axis of $\sim19.4$~AU, the separation of $\iota$~Dra and the outer companion is $\sim630$~mas. Assuming a radius of 1~$R_J$ and albedo of 0.5, the peak brightness of the companion is estimated to be $\sim2 \times 10^{-10}$. With a predicted contrast ratio detection threshold of $\sim10^{-10}$, the Habitable Exoplanet Observatory (HabEx) combined with starshade \citep{reporthabex} will have the greatest ability to directly image the outer companion of $\iota$~Dra, provided noise estimation for the instrument is accurate. Other missions, such as the Nancy Grace Roman Space Telescope with no star shade and a limiting contrast ratio of $\sim1e^{-9}$ \citep{Roman2021}, are unlikely to able to detect the companion.


\section{Conclusions}
\label{conclusions}

Planets orbiting giant stars are fascinating systems that provide opportunities to examine the effects of stellar evolution on the dynamics of planetary orbits. The case of $\iota$~Dra provides a relatively nearby multi-planet giant-star system through which to study these effects through precise measurements of the stellar and planetary properties. Through our SED and asteroseismology analysis we refined the stellar parameters for $\iota$~Dra. Using {\it TESS} observations of the star over 5 sectors, we were able to constrain the stellar radius to $\sim$2\,\%, mass to $\sim$6\,\%, and age to $\sim$28\,\%. Investigation into the evolutionary state of the star points towards an RGB classification. 

We obtained 456 new RV observations of $\iota$~Dra with the Lick Observatory APF telescope between February 2018 to February 2021. These RV observations, combined with those previously published, cover several orbital periods of the known planet, providing significant improvement to the orbital parameters. These in turn allowed us to detect curvature in the previously identified RV linear trend which is likely caused by a previously undiscovered outer companion. After running a dynamical analysis with the \texttt{MEGNO} chaos indicator to determine the range of semi-major axis and eccentricity configurations that the orbiting body could exist within, we further constrained the possible orbits with \texttt{The Joker}. We then combined our RV data with astrometry from Gaia and Hippacos with the help of \texttt{orvara} and were able to obtain a best-fit solution for the outer companion. This solution gave the outer companion a period of $68^{+60}_{-36}$ years, and eccentricity of $0.455^{+0.12}_{-0.084}$.

The \texttt{orvara} preferred orbital solution for the sub-stellar outer companion estimated a mass on the border of the planet and brown dwarf regimes. The exclusion of stellar mass companions from the speckle imaging presented in \citep{kane2014c} also suggests that the orbiting companion is sub-stellar. As it is ambiguous as to whether the companion is burning deuterium, we are unable to confidently identify whether it is a planet or a brown dwarf. As brown dwarfs are relatively rare, with {0.8}$_{-0.5}^{+0.8}$\% of stars hosting a brown dwarf \citep{nielsen2019c}, this could be an important addition to the population of brown dwarfs. The relatively close proximity of $\iota$~Dra to Earth will make this a prime target in future giant planet and brown dwarf characterization studies. Continued observations of this target will help refine the orbital parameters of the outer companion and confirm its classification as either a giant planet or a brown dwarf.


\section*{Acknowledgements}

The results reported herein benefited from collaborations and/or information exchange within NASA's Nexus for Exoplanet System Science (NExSS) research coordination network sponsored by NASA's Science Mission Directorate. P.D. acknowledges support from a National Science Foundation Astronomy and Astrophysics Postdoctoral Fellowship under award AST-1903811. T.L.C.~acknowledges support from the European Union's Horizon 2020 research and innovation programme under the Marie Sk\l{}odowska-Curie grant agreement No.~792848 (PULSATION). T.L.C.~is supported by Funda\c c\~ao para a Ci\^encia e a Tecnologia (FCT) in the form of a work contract (CEECIND/00476/2018).
C.K. acknowledges support by Erciyes University Scientific Research Projects Coordination Unit under grant number MAP-2020-9749.
T.L. acknowledges the funding from the European Research Council (ERC) under the European Union’s Horizon 2020 research and innovation programme (CartographY GA. 804752).
D.B. and M.S.C. acknowledge supported by FCT through the research grants UIDB/04434/2020, UIDP/04434/2020 and PTDC/FIS-AST/30389/2017, and by FEDER - Fundo Europeu de Desenvolvimento Regional through COMPETE2020 -
Programa Operacional Competitividade e Internacionalização (grant: POCI-01-0145-FEDER-030389). M.S.C. is supported by national funds through FCT in the form of a work contract. R.A.G. and S.N.B. acknowledge the support of the PLATO and GOLF CNES grants. S.M. acknowledges the support from the Spanish Ministry of Science and Innovation with the Ramon y Cajal fellowship number RYC-2015-17697 and with the grant number PID2019-107187GB-I00.
D.H. acknowledges support from the Alfred P. Sloan Foundation, the National Aeronautics and Space Administration (80NSSC19K0379), and the National Science Foundation (AST-1717000). D.L.B. acknowledges support from the NASA TESS GI Program under awards 80NSSC18K1585 and 80NSSC19K0385. T.R.B. acknowledges support from the Australian Research Council (DP210103119).


\software{
\texttt{RadVel} \citep[\url{https://github.com/California-Planet-Search/radvel}]{fulton2018a},
\texttt{lightkurve} \citep[\url{https://github.com/hvidy/tessbkgd/blob/stable/notebooks/iot_Dra_tpf.ipynb}]{lightkurve} \texttt{DIAMONDS} \citep[\url{https://github.com/EnricoCorsaro/DIAMONDS};][]{DIAMONDS}, Background extension to \texttt{DIAMONDS} \citep[\url{https://github.com/EnricoCorsaro/Background};][]{DIAMONDS}, \texttt{FAMED} \citep[\url{https://github.com/EnricoCorsaro/FAMED};][]{famed},
\texttt{REBOUND: MEGNO} \citep[\url{https://github.com/hannorein/rebound}]{rein2012a},
\texttt{The Joker} \citep[\url{https://github.com/adrn/thejoker};][]{Price_Whelan_2017}
}, \texttt{htof} \citep[\url{https://github.com/gmbrandt/htof}][], \texttt{orvara} \citep[\url{https://github.com/t-brandt/orvara};][]{orvara2021}

\clearpage
\newpage

\appendix
\label{APPRV}

\section{Individual Mode Frequencies}
\label{APPFreq}

Table \ref{tab:modes} lists all significant modes (i.e., $p_{\rm det} \ge 0.993$) returned by \texttt{FAMED}. $p_{\rm det}$ is the peak detection probability based on a Bayesian model comparison as performed by \texttt{FAMED}. A peak is tested against the noise only if its height in the smoothed power spectrum is lower than 10 times the local level of the background, otherwise the peak is automatically considered as detected (denoted as `---'). A peak is deemed significant if $p_{\rm det} \ge 0.993$. See sect.~5.3 of \citet{famed} for details. The List column indicates which list each mode belongs to: Min. = Belongs to Minimal List; Max. = Belongs to Maximal List (but not to Minimal List).

\begin{deluxetable}{ccccc}
\tablecolumns{5}
\tablewidth{0pc}
\tablecaption{\label{tab:modes} Observed mode frequencies.}
\tablehead{
\colhead{$\ell$} & \colhead{Frequency ($\rm{\mu Hz}$)} & \colhead{1-$\sigma$ Uncertainty ($\rm{\mu Hz}$)} & \colhead{$p_{\rm det}$\tablenotemark{a}
} & \colhead{List\tablenotemark{b}
}}
\startdata
1&30.384&0.028&0.997&Min. \\
2&31.538&0.133&1.000&Min. \\
0&32.024&0.024&0.998&Min. \\
1&33.913&0.088&1.000&Min. \\
2&35.410&0.039&---&Min. \\
0&35.878&0.035&---&Min. \\
1&37.983&0.035&---&Min. \\
2&39.361&0.072&---&Min. \\
0&39.904&0.049&---&Min. \\
3&40.684&0.056&1.000&Max. \\
1&42.078&0.024&---&Max. \\
1&42.552&0.016&---&Min. \\
1&43.063&0.011&---&Max. \\
2&43.530&0.107&---&Min. \\
0&43.925&0.016&---&Max. \\
1&45.980&0.027&---&Min. \\
2&47.364&0.072&---&Min. \\
0&48.015&0.038&---&Min. \\
1&49.948&0.027&---&Min. \\
1&50.420&0.022&0.997&Max. \\
1&54.274&0.021&0.999&Min. \\
2&55.202&0.100&0.999&Max. \\
0&55.565&0.025&0.994&Max. \\
\enddata
\tablenotetext{a}{\small Peak detection probability based on a Bayesian model comparison as performed by \texttt{FAMED}. A peak is tested against the noise only if its height in the smoothed power spectrum is lower than 10 times the local level of the background, otherwise the peak is automatically considered as detected (denoted as `---'). A peak is deemed significant if $p_{\rm det} \ge 0.993$. See sect.~5.3 of \citet{famed} for details.}
\tablenotetext{b}{\small Min. = Belongs to Minimal List; Max. = Belongs to Maximal List (but not to Minimal List).}
\end{deluxetable}

\section{Stellar Model Grid Description}
\label{APPGrid}

We used Modules for Experiments in Stellar Astrophysics ({\sc mesa}, version 12115) to construct a grid of stellar models. General descriptions of the input physics and numerical methods can be found in the {\sc mesa} papers \citep{paxton2011,paxton2013, paxton2015,paxton2018}. We adopted the solar chemical mixture [$(Z/X)_{\odot}$ = 0.0181] provided by \citet{asplundetal2009}. We used the {\sc mesa} $\rho$--$T$ tables based on the 2005 update of the OPAL equation of state tables \citep{rogers&nayfonov2002} and we used OPAL opacities supplemented by the low-temperature opacities from \citet{fergusonetal2005}. The {\sc mesa} ‘simple’ photosphere was used for the set of boundary conditions for modeling the atmosphere; alternative model atmosphere choices do not strongly affect the results for solar-like oscillators \citep{yildiz2007,joyce&chaboyer2018b,nsambaetal2018,vianietal2018} or for $\delta$\,Sct stars (Murphy et al., in review). The mixing-length theory of convection was implemented, where $\alpha_{\rm MLT} = \ell_{\rm MLT}/H_p$ is the mixing-length parameter. The exponential scheme by \citet{herwig2000} was adopted for the convective overshooting. We defined the overshooting parameter as $f_{\rm{ov}} = (0.13\,M_\ast - 0.098)/9.0$  and adopted a fixed $f_{\rm{ov}}$ of 0.018 for models with a mass above $2.0\,{\rm M}_{\odot}$, following the mass-overshooting relation found by \citet{2010ApJ...718.1378M}. We also applied the {\sc mesa} predictive mixing scheme in our model for a smooth convective boundary. The mass loss rate on the red-giant branch follows the Reimers' prescription with $\eta = 0.2$, which is constrained by the old open clusters NGC\,6791 and NGC\,6819 \citep{miglioetal2012}. Models in the grid varied in stellar mass within 0.8 -- $2.2\,{\rm M}_{\odot}$ in steps of $0.02\,{\rm M}_{\odot}$, in initial helium fraction ($Y_{\rm init}$) within 0.24 -- 0.32 in steps of 0.02, and in initial metallicity ([Fe/H]) within $-0.5$ -- 0.5 in steps of 0.1. Moreover, four values were considered for the mixing length parameter associated with the description of convection, namely,  $\alpha_{\rm MLT}$ = 1.7, 1.9, 2.1, and 2.3.


\end{document}